%% file: tanimoto.tex
\documentclass[times]{aastex631}
\usepackage{amsmath}
\usepackage{hyperref}

\begin{document}
\title{Circumnuclear Multi-phase Gas in the Circinus Galaxy. V. The Origin of the X-Ray Polarization in the Circinus Galaxy}
\correspondingauthor{Atsushi Tanimoto}
\email{atsushi.tanimoto@sci.kagoshima-u.ac.jp}
\author[0000-0002-0114-5581]{Atsushi Tanimoto}
\affiliation{Graduate School of Science and Engineering, Kagoshima University, Kagoshima 890-0065, Japan}
\author[0000-0002-8779-8486]{Keiichi Wada}
\affiliation{Graduate School of Science and Engineering, Kagoshima University, Kagoshima 890-0065, Japan}
\affiliation{Research Center for Space and Cosmic Evolution, Ehime University, Matsuyama 790-8577, Japan}
\affiliation{Faculty of Science, Hokkaido University, Sapporo 060-0810, Japan}
\author[0000-0003-0548-1766]{Yuki Kudoh}
\affiliation{Graduate School of Science and Engineering, Kagoshima University, Kagoshima 890-0065, Japan}
\affiliation{Astronomical Institute, Tohoku University, Miyagi 980-8578, Japan}
\author[0000-0003-2670-6936]{Hirokazu Odaka}
\affiliation{Department of Earth and Space Science, Osaka University, Osaka 560-0043, Japan}
\author[0000-0001-6653-779X]{Ryosuke Uematsu}
\affiliation{Department of Astronomy, Kyoto University, Kyoto 606-8502, Japan}
\author[0000-0002-5701-0811]{Shoji Ogawa}
\affiliation{Department of Astronomy, Kyoto University, Kyoto 606-8502, Japan}
\affiliation{Institute of Space and Astronautical Science, Japan Aerospace Exploration Agency, Kanagawa 252-5210, Japan}

\input{Abstract}
\input{Section0101}
\input{Section0201}
\input{Figure01}
\input{Figure02}
\input{Figure03}
\input{Section0202}
\input{Figure04}
\input{Section0301}
\input{Figure05}
\input{Figure06}
\input{Section0302}
\input{Figure07}
\input{Figure08}
\input{Section0401}
\input{Section0402}
\input{Section0501}
\input{Acknowledgment}

\bibliography{tanimoto}
\bibliographystyle{aasjournal}
\end{document}

%% file: Abstract.tex
\begin{abstract}
The Imaging X-ray Polarimetry Explorer (IXPE) detected X-ray polarization in the nearest Seyfert 2 galaxy, the Circinus galaxy, for the first time. To reproduce the IXPE results, we computed the degree of polarization based on two types of radiative hydrodynamic simulations: a parsec-scale three-dimensional model and a sub-parsec-scale axisymmetric model with a higher spatial resolution. In a series of papers, we confirmed that these models naturally explain the multi-wavelength observations of the Circinus galaxy from radio to X-rays. We used a Monte Carlo Simulation for Astrophysics and Cosmology code to compute the linear polarization of continuum emission. We found that the degree of polarization based on the parsec-scale radiation-driven fountain model was smaller than that observed with the IXPE. The degree of polarization based on the sub-parsec-scale model depends on the hydrogen number density of the disk ($d$), and the degree of polarization obtained from our simulation is consistent with that observed with the IXPE in the case of $\log d/\mathrm{cm}^{-3} \geq 13$. We investigate where the photons are Compton scattered and imply that the origin of the X-ray polarization in the Circinus galaxy is the outflow inside $0.01 \ \mathrm{pc}$. In this case, the degree of polarization may change over a timescale of approximately ten years. 
\end{abstract}.
\keywords{Active galactic nuclei (16), Astrophysical black holes (98), High energy astrophysics (739), Seyfert galaxies (1447), Supermassive black holes (1663), X-ray active galactic nuclei (2035)}

%% file: Section0101.tex
\section{Introduction}\label{Section0101}
A unified model of active galactic nuclei (AGNs) suggests that a structure composed of gas and dust, called a torus, surrounds accreting supermassive black holes (SMBHs) \citep{Antonucci1993, Urry1995, Netzer2015, Ramos2017}. This torus plays an essential role in the co-evolution between SMBHs and their host galaxies \citep[e.g.,][]{Kormendy2013}. This is because the torus is believed to exist between the SMBH and the host galaxy, providing a mass from the host galaxy to the SMBH. However, the mechanism of torus formation and the structure of the sub-parsec-scale gas remain unclear.

X-ray observations are among the most powerful tools for probing the structures of torus-scale gases. In particular, the polarization of X-rays caused by Compton scattering can provide clues to the geometry of the torus, such as the hydrogen column density along the equatorial direction ($\log N_{\mathrm{H}}^{\mathrm{Equ}}/\mathrm{cm}^{-2}$) and the opening angle of the torus ($\theta_{\mathrm{open}}$). In fact, several studies have investigated the dependence of the degree of polarization on torus properties \citep{Goosmann2011, Marin2016, Marin2018, Ratheesh2021}. For instance, \cite{Ratheesh2021} assumed a partially spherically symmetric torus and studied the dependence of the degree of polarization on $\log N_{\mathrm{H}}^{\mathrm{Equ}}/\mathrm{cm}^{-2}$ and $\theta_{\mathrm{open}}$. \cite{Ratheesh2021} found that the degree of polarization increased with $\log N_{\mathrm{H}}^{\mathrm{Equ}}/\mathrm{cm}^{-2}$ \citep[][Figure~10]{Ratheesh2021}.

Recently, the Imaging X-ray Polarimetry Explorer (IXPE; \citealt{Weisskopf2022}) measured the degree of polarization of the Circinus galaxy for the first time, one of the nearest ($4.2 \pm 0.8 \ \mathrm{Mpc}$: \citealt{Freeman1977}) Compton-thick AGNs whose hydrogen column density along the line of sight is greater than $10^{24} \ \mathrm{cm}^{-2}$ \citep{Arevalo2014, Tanimoto2019, Uematsu2021}. \cite{Ursini2023} has analyzed the observational data and indicated that the degree of polarization is $28 \pm 7\%$ for the neutral reflector. To explain the degree of polarization of the Circinus galaxy, \cite{Ursini2023} calculated it based on the simple torus model with the same geometric structure as that reported by \cite{Ratheesh2021}. As a result, they were able to reproduce the degree of polarization of the Circinus galaxy with $\log N_{\mathrm{H}}^{\mathrm{Equ}}/\mathrm{cm}^{-2} = 25$ and $\theta_{\mathrm{open}} \simeq 50\degr$. However, in the case of the simple model, any geometric structure can be assumed to reproduce the observational data. Since the real structures of the central region of AGNs are still unknown, it is essential for us to study the origin of X-ray polarization based on radiative hydrodynamical simulations.

In the fifth paper in this series, we computed the degree of polarization based on radiative hydrodynamic simulations (Paper I: \citealt{Wada2018a}, Paper I\hspace{-1.2pt}I: \citealt{Izumi2018}, Paper I\hspace{-1.2pt}I\hspace{-1.2pt}I: \citealt{Wada2018b}, and Paper I\hspace{-1.2pt}V: \citealt{Uzuo2021}). Here, we used two models: a parsec-scale three-dimensional model \citep{Wada2012, Wada2015} and a sub-parsec-scale axisymmetric model \citep{Kudoh2023}. \cite{Wada2012} considered the radiative feedback from the central source such as the radiation pressure on the dusty gas and the X-ray heating and showed that a geometrically and optically thick torus-like structure could be naturally formed based on the three-dimensional radiative hydrodynamic simulation. In fact, their model successfully reproduced the molecular and atomic lines in the radio spectrum \citep{Wada2018a, Izumi2018, Uzuo2021}, the infrared spectral energy density \citep{Wada2016}, the properties of the narrow line regions \citep{Wada2018b}, and the X-ray spectrum observed with the Chandra/High Energy Transmission Grating \citep{Ogawa2022} of the Circinus galaxy. However, \cite{Buchner2021} computed the X-ray spectra of neutral materials based on a parsec-scale radiation-driven fountain model, and suggested that their model could not explain the Compton hump in the Circinus galaxy. To reproduce the Compton hump, they reported that a Compton-thick material on the sub-parsec scale is required. If such a material exists on the sub-parsec scale, it affects the degree of polarization. Therefore, we computed the degree of polarization based on the sub-parsec-scale radiation-driven fountain model \citep{Kudoh2023}. The sub-parsec-scale radiation-driven fountain model has identified time variability in the Balmer lines over a scale of 1--10 years \citep{Wada2023}. The X-ray spectra also exhibit time variability in some AGNs \citep{Ricci2022}. Hence, it would be interesting to investigate whether the degree of polarization displays similar variability and if the radiation-driven fountain contributes to it.

The remainder of this paper is organized as follows: In \hyperref[Section0200]{Section~2}, we presented the Monte Carlo simulations and the radiation-driven fountain model. In \hyperref[Section0300]{Section~3}, we described the results of our simulations. In \hyperref[Section0400]{Section~4}, we discussed the origin of the X-ray polarization in the Circinus galaxy and the time variability of the degree of polarization. Finally, in \hyperref[Section0500]{Section~5}, we presented our conclusions.

%% file: Section0201.tex
\section{Methods}\label{Section0200}
\subsection{Stokes Parameters and Monte Carlo Simulations}\label{Section0201}
To describe the polarization state, we defined the Stokes parameters. The Stokes parameters $I$, $Q$, $U$, and $V$ are defined using the Poincare sphere as follows:
\begin{align}
Q   & = IP \cos{2\chi} \cos{2\psi}\\
U   & = IP \cos{2\chi} \sin{2\psi}\\
V   & = IP \sin{2\chi}.
\end{align}
Here, $P$ is the degree of polarization, $2\chi$ is the elevation angle of the Poincare sphere, and $2\psi$ is the azimuth angle of the Poincare sphere. As the IXPE can only detect linear polarization, we assumed that $\chi$ is equal to $0$. In this case, the Stokes parameters $I$, $Q$, and $U$ are defined as follows:
\begin{align}
Q   & = IP \cos{2\psi}\\
U   & = IP \sin{2\psi}.               
\end{align}
From equations (4) and (5), the degree of polarization $P$ can be expressed as,
\begin{equation}
P   = \frac{\sqrt{Q^2+U^2}}{I}.
\end{equation}

To compute the Stokes parameters, we used the Monte Carlo Simulation for Astrophysics and Cosmology (MONACO; \citealt{Odaka2011, Odaka2016}) code version 1.7.s1. This code uses the Geant4 code \citep{Agostinelli2003, Allison2006, Allison2016} to track photons in complex geometric structures. Although the Geant4 code implements physical processes, MONACO uses physical processes optimized for astrophysics. Currently, MONACO has three sets of physical processes: (1) X-ray reflection from neutral matter \citep{Odaka2011, Furui2016, Tanimoto2019, Tanimoto2022, Uematsu2021}, (2) Comptonization in a hot flow \citep{Odaka2013, Odaka2014}, and (3) photon interactions in a photoionized plasma \citep{Watanabe2006, Hagino2015, Hagino2016, Tomaru2018, Tomaru2020, Mizumoto2021}.

For simplicity, we used the X-ray reflection from neutral matter in this study. In other words, we assumed that the thermal motion of the gas and ionized matter was ignored. Three physical processes were considered: photoelectric absorption, fluorescence line emission, and Compton scattering caused by free electrons. We assumed that the primary photons were unpolarized. Furthermore, we considered the solar abundances from \cite{Anders1989}.

%% file: Figure01.tex
\begin{figure}\label{Figure01}
\gridline{\fig{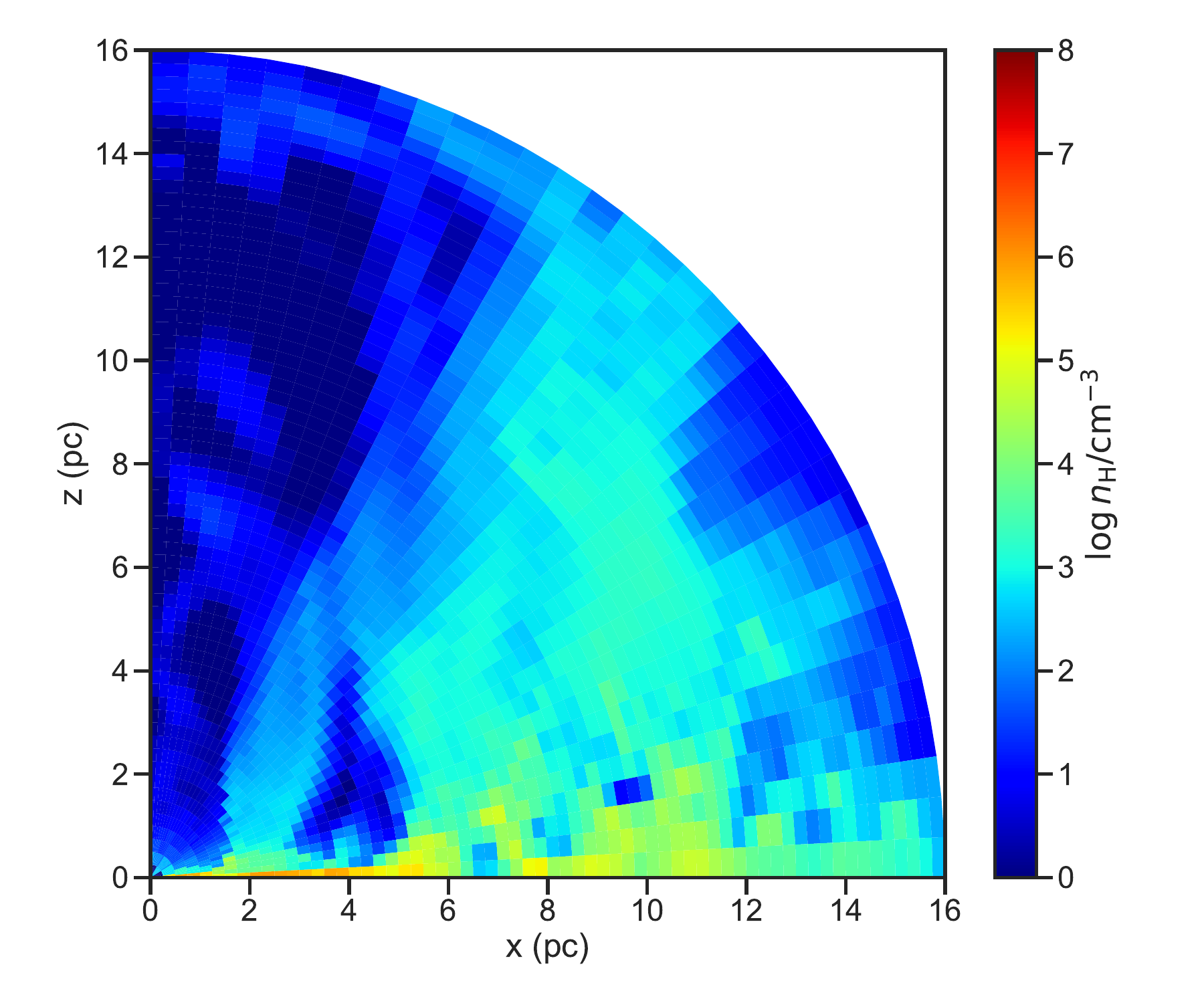}{0.5\textwidth}{(a)}\fig{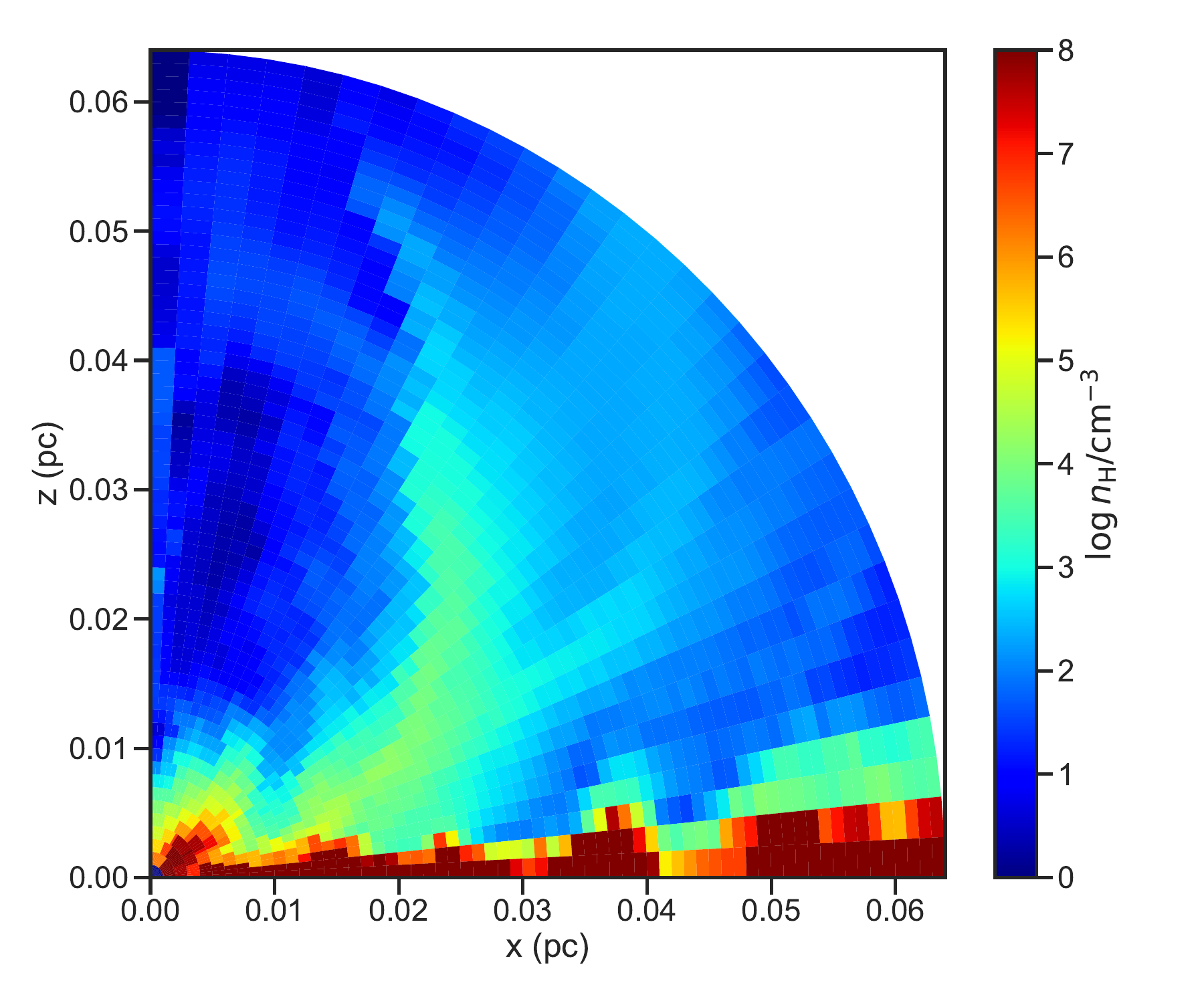}{0.5\textwidth}{(b)}}
\caption{(a) The distribution of the hydrogen number density ($\log n_{\mathrm{H}}/\mathrm{cm}^{-3}$) based on the parsec-scale radiation-driven fountain model \citep{Wada2016}. (b) The distribution of $\log n_{\mathrm{H}}/\mathrm{cm}^{-3}$ based on the sub-parsec-scale radiation-driven fountain model \citep{Kudoh2023}. Here we assumed that the logarithmic initial disk density was $\log d/\mathrm{cm}^{-3} = 13$.}
\end{figure}

%% file: Figure02.tex
\begin{figure}\label{Figure02}
\gridline{\fig{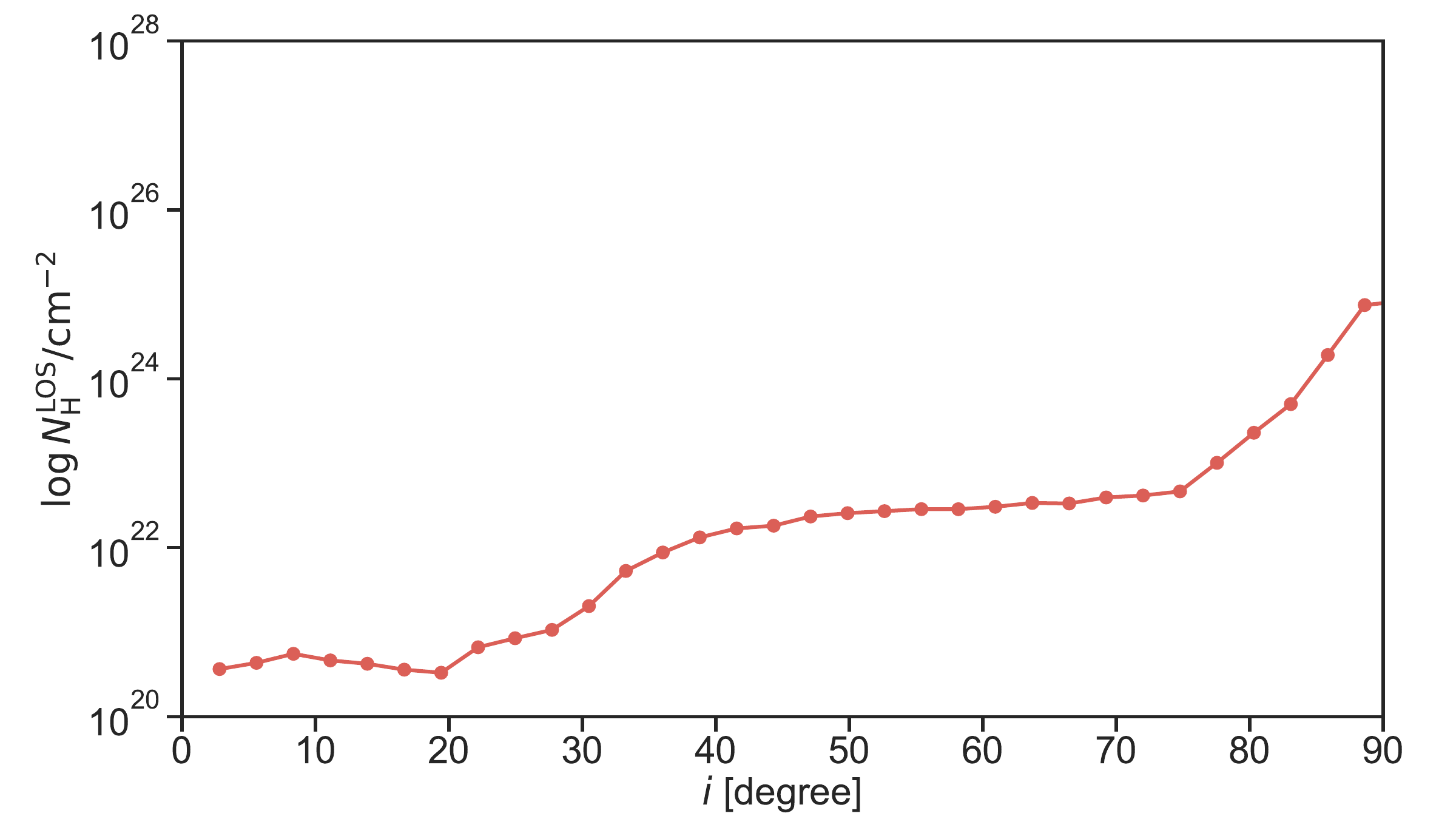}{0.5\textwidth}{(a)}\fig{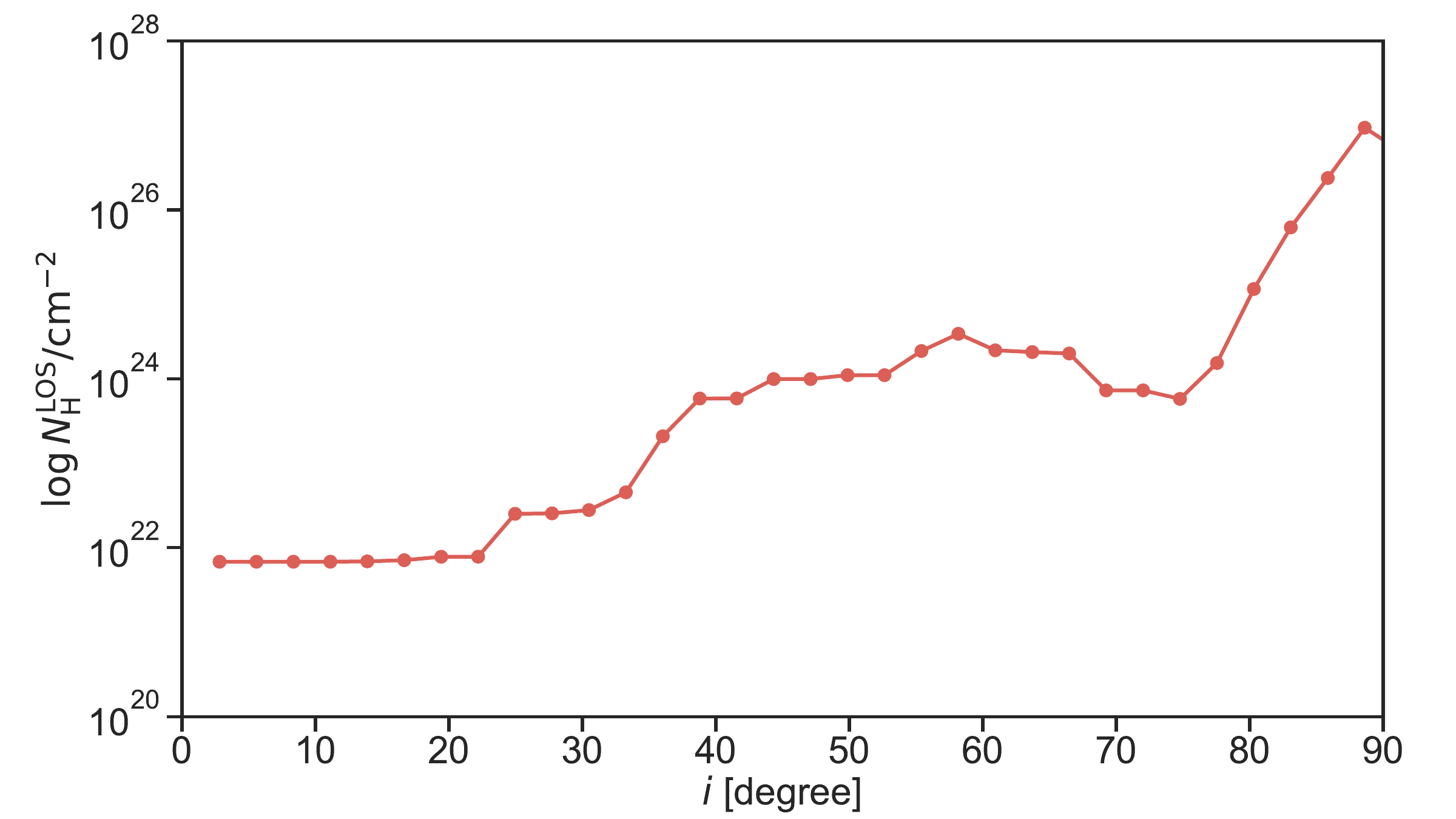}{0.5\textwidth}{(b)}}
\caption{(a) The hydrogen column density along the line of sight ($\log N_{\mathrm{H}}^{\mathrm{LOS}}/\mathrm{cm}^{-2}$) as a function of the inclination angle ($i$) based on the parsec-scale radiation-driven fountain model. (b) $\log N_{\mathrm{H}}^{\mathrm{LOS}}/\mathrm{cm}^{-2}$ as the function of $i$ based on the sub-parsec-scale radiation-driven fountain model. Here we assumed that the logarithmic initial disk density was $\log d/\mathrm{cm}^{-3} = 13$.}
\end{figure}

%% file: Figure03.tex
\begin{figure}\label{Figure03}
\gridline{\fig{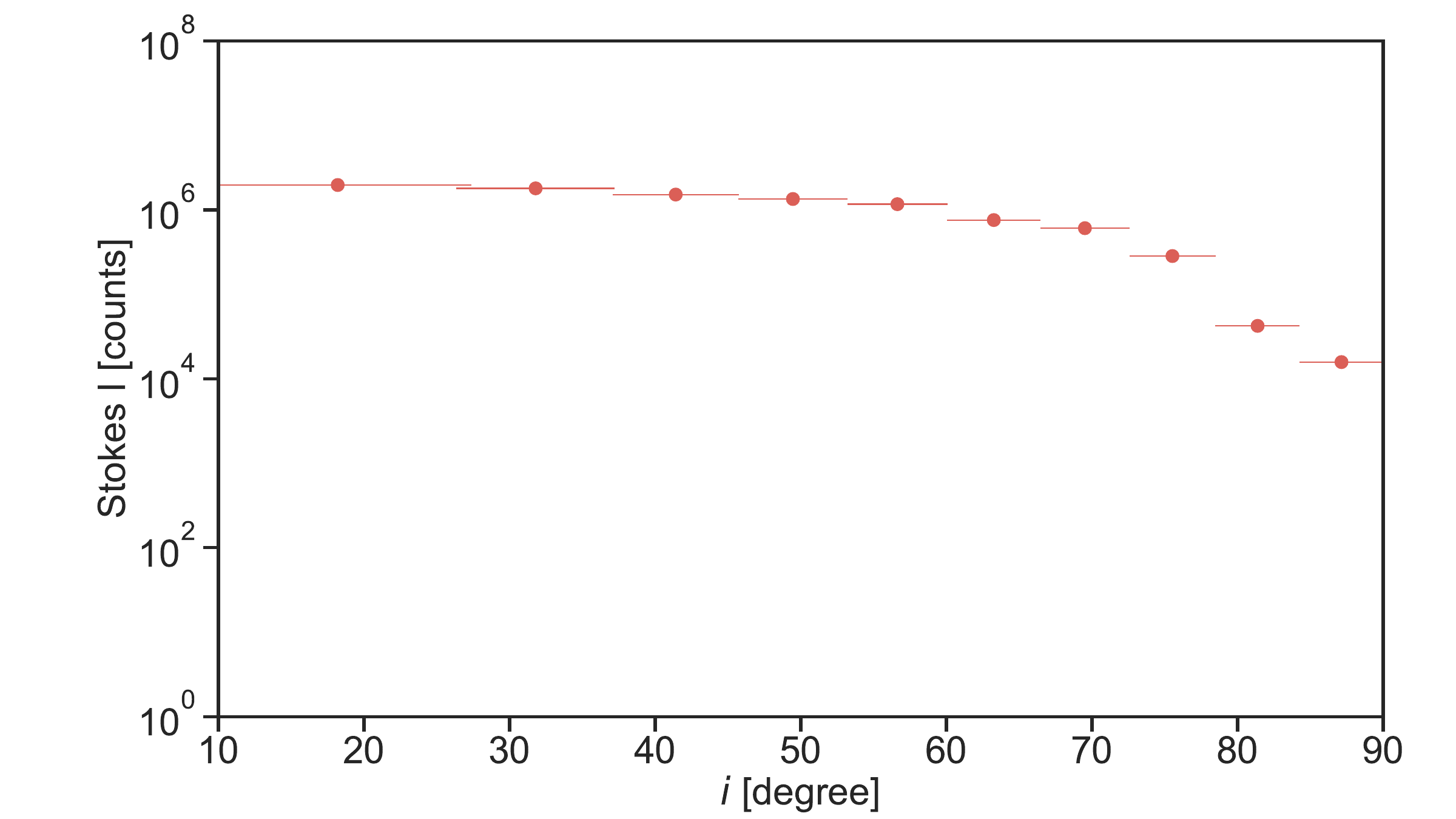}{0.5\textwidth}{(a)}\fig{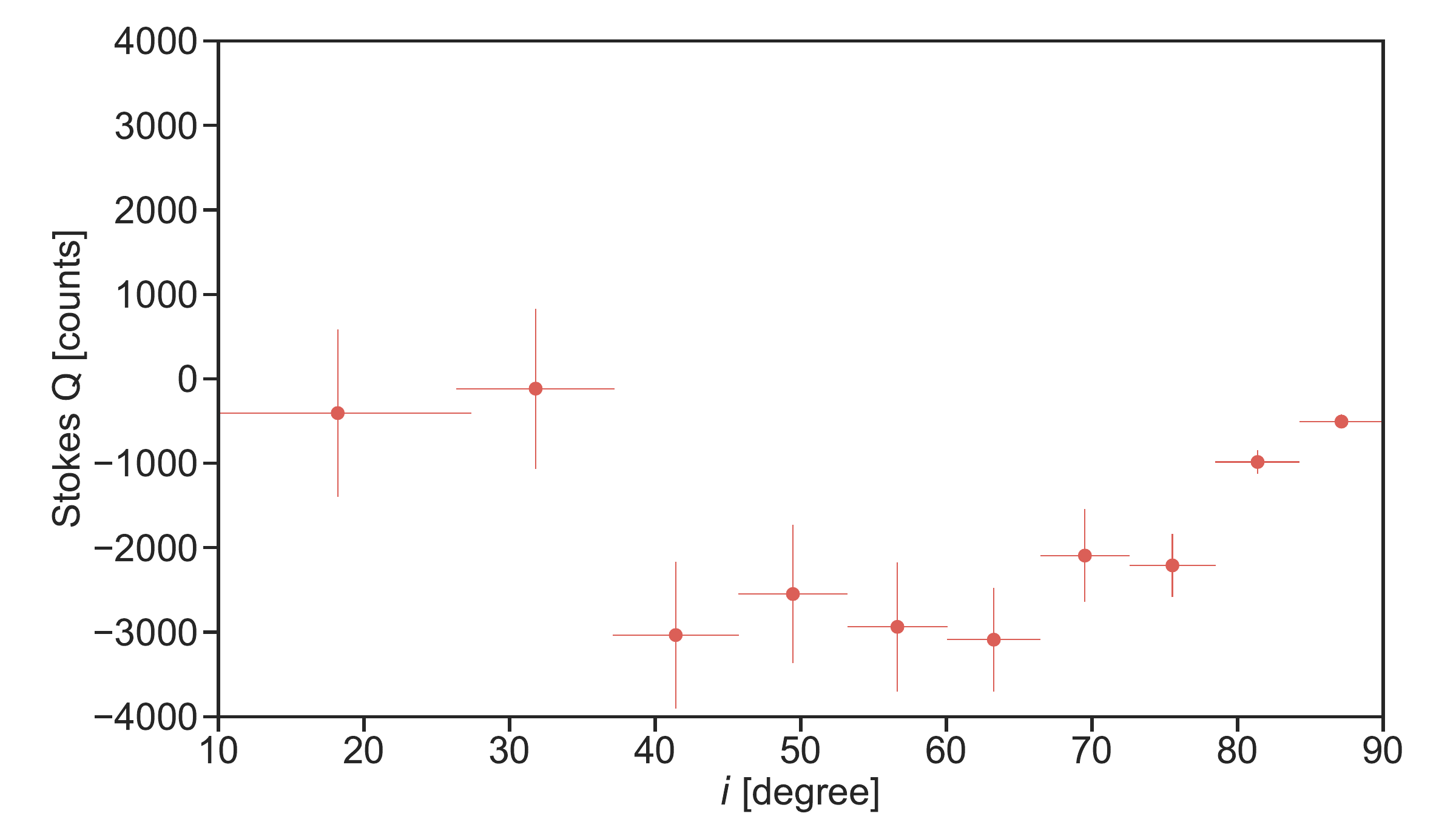}{0.5\textwidth}{(b)}}
\gridline{\fig{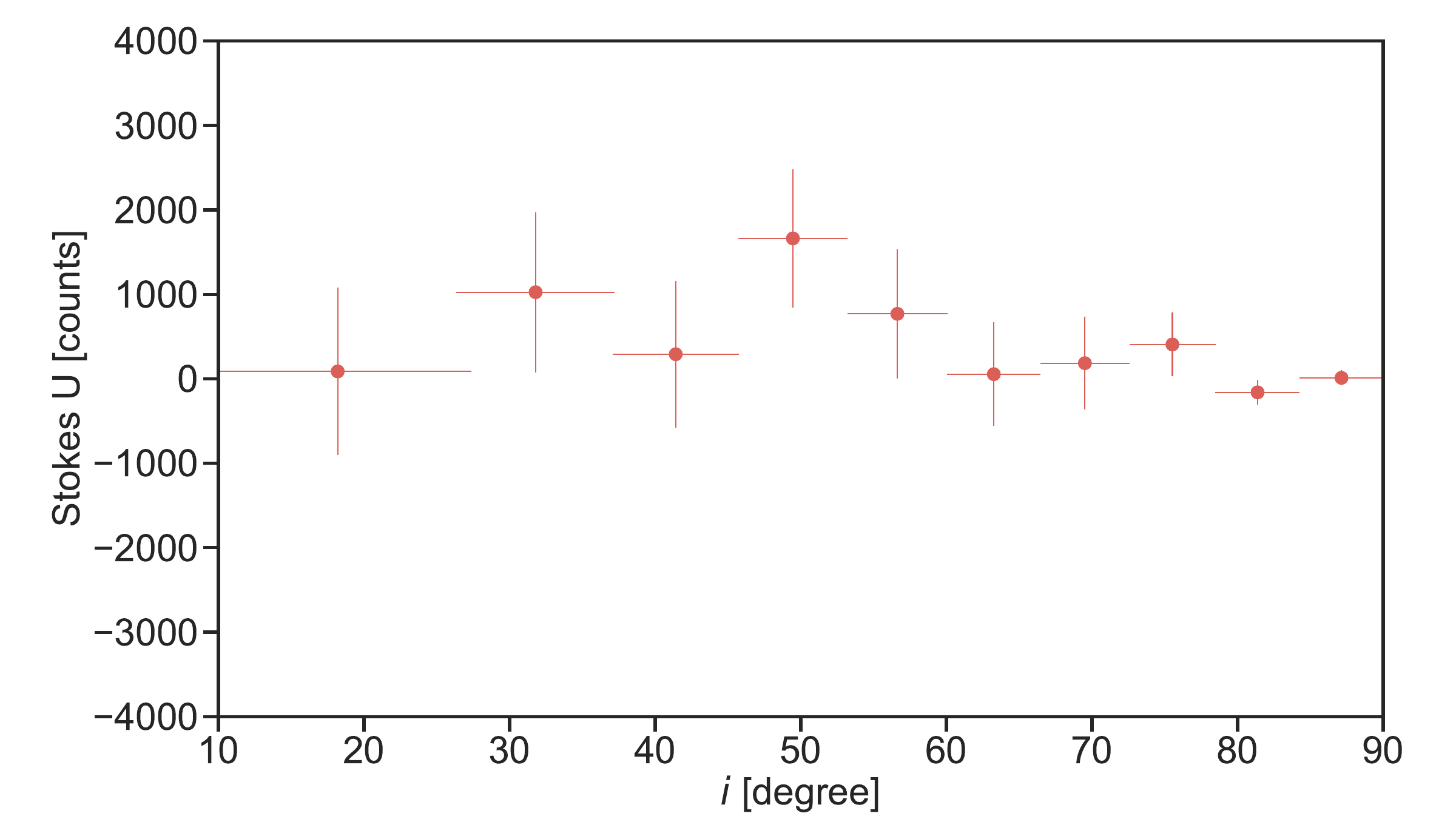}{0.5\textwidth}{(c)}\fig{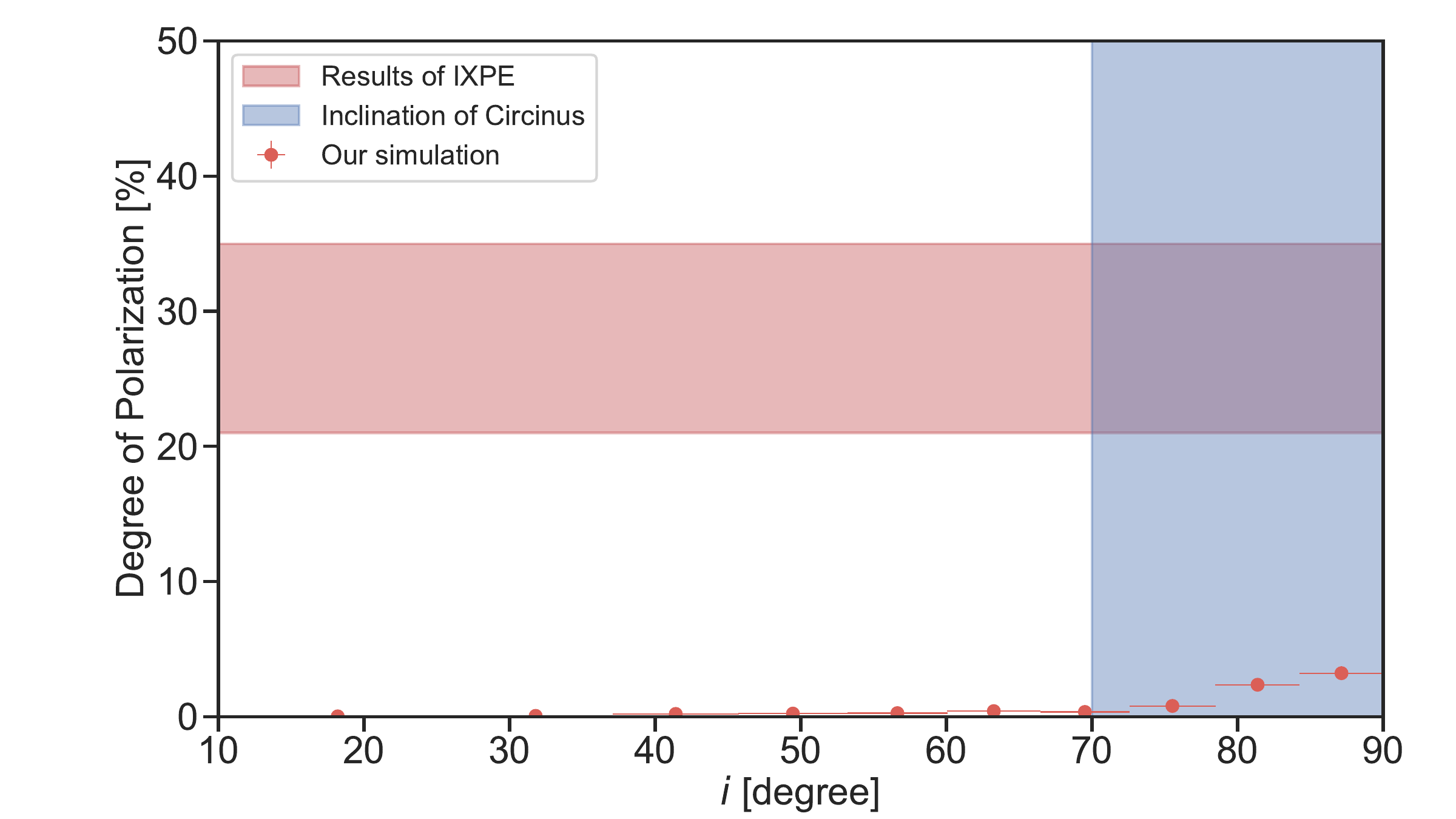}{0.5\textwidth}{(d)}}
\caption{(a) The dependence of the Stokes parameter $I$ on the inclination angle ($i$). (b) The dependence of the Stokes parameter $Q$ on $i$. (c) The dependence of the Stokes parameter $U$ on $i$. (d) Comparison of the degree of polarization obtained from our simulation with that of the Circinus galaxy observed with the IXPE (red region). The blue region represents a range of inclination angles of the Circinus galaxy suggested by observations (see \hyperref[Section0301]{Section~3.1}). Here errors correspond to a 1$\sigma$ credible interval.}
\end{figure}

%% file: Section0202.tex
\subsection{Radiation-Hydrodynamic Models}\label{Section0202}
First, we computed the degree of polarization based on the parsec-scale radiation-driven fountain model \citep{Wada2016} (\hyperref[Section0301]{Section~3.1}). \cite{Wada2016} computed the three-dimensional radiative dynamical simulation ranging from $-16 \ \mathrm{pc}$ to $+16 \ \mathrm{pc}$ in the $256^3$ Cartesian coordinate system. The model consistently reproduced the observed spectral energy density of the Circinus galaxy when the inclination angle was greater than $75\degr$. They assumed that the black hole mass was $2.0 \times 10^{6} M_{\sun}$ based on water maser observations \citep{Greenhill2003}, which is also consistent with the molecular gas kinematics revealed by ALMA \citep{Izumi2023}, the Eddington ratio was $20\%$ \citep{Arevalo2014}, and the X-ray luminosity from 2 keV to 10 keV was $2.8 \times 10^{42} \ \mathrm{erg} \ \mathrm{s}^{-1}$ \citep{Arevalo2014} of the Circinus galaxy. We used the same input data analyzed by \cite{Ogawa2022}, which reformed the $-16 \ \mathrm{pc}$ to $+16 \ \mathrm{pc}$ range from a $256^3$ Cartesian coordinate grid to a $64^3$ spherical coordinate grid. \hyperref[Figure01]{Figure~1(a)} shows the distribution of the hydrogen number density based on the parsec-scale radiation-driven fountain model and \hyperref[Figure02]{Figure~2(a)} shows the hydrogen column density along the line of sight as a function of the inclination angle. \hyperref[Figure02]{Figure~2(a)} shows that the parsec-scale radiation-driven fountain model is Compton-thin for almost all the inclination angles. We generated $10^{8}$ photons in the 1--10 keV range with a photon index of 1.8 of the Circinus galaxy \citep{Tanimoto2019}.

The second hydrodynamic model is the sub-parsec-scale radiation-driven fountain model \citep{Kudoh2023} (\hyperref[Section0302]{Section~3.2}). \cite{Kudoh2023} computed the radiation-driven fountain model considering the more inner regions based on the two-dimensional radiative hydrodynamic simulations using the coordinated astronomical numerical software plus (CANS+: \citealt{Matsumoto2019}). To apply this model to the Circinus galaxy, we assumed the same black hole mass, Eddington ratio, and X-ray luminosity as in \cite{Wada2016}. This model has the initial hydrogen number density of disk $\log d/\mathrm{cm}^{-3}$ as a free parameter. In this study, we studied four cases of $\log d/\mathrm{cm}^{-3}$: $\log d/\mathrm{cm}^{-3} = 12, 13, 14, \ \mathrm{and} \ 15$. This is because, if $\log d/\mathrm{cm}^{-3}$ is less than 11, we cannot reproduce the observational results of the IXPE. On the other hand, an excessively large $\log d/\mathrm{cm}^{-3}$ is unrealistic because it should be dynamically unstable. We performed two-dimensional axisymmetric radiative hydrodynamic simulations over a region inside $0.256 \ \mathrm{pc}$ divided into $256$ grids in the radial direction and a region from $-0.256 \ \mathrm{pc}$ to $+0.256 \ \mathrm{pc}$ divided into $512$ grids in the z-axis direction. Since the outer regions have a small density and do not affect the results, we used simulation data inside $0.064 \ \mathrm{pc}$ regions divided into $64^3$ grids in spherical coordinates. \hyperref[Figure01]{Figure~1(b)} shows the distribution of the hydrogen number density based on the sub-parsec-scale radiation-driven fountain model and \hyperref[Figure02]{Figure~2(b)} shows the hydrogen column density along the line of sight as the function of the inclination angle. We found that the sub-parsec-scale radiation-driven fountain model had a Compton-thick material when $i$ was greater than $\simeq 50\degr$. To study the dependence of the degree of polarization on $\log d/\mathrm{cm}^{-3}$ and time, we generated $10^{8}$ photons in the range of 1--10 keV, with a photon index of 1.8 for each of the four patterns of $\log d/\mathrm{cm}^{-3}$ and four patterns of snapshots. Since we do not know how the Circinus galaxy is tilted with respect to the celestial north pole, we assumed coordinates where the z-axis direction of the radiative hydrodynamics simulation is negative in Stokes parameter $Q$.

%% file: Figure04.tex
\begin{figure}\label{Figure04}
\gridline{\fig{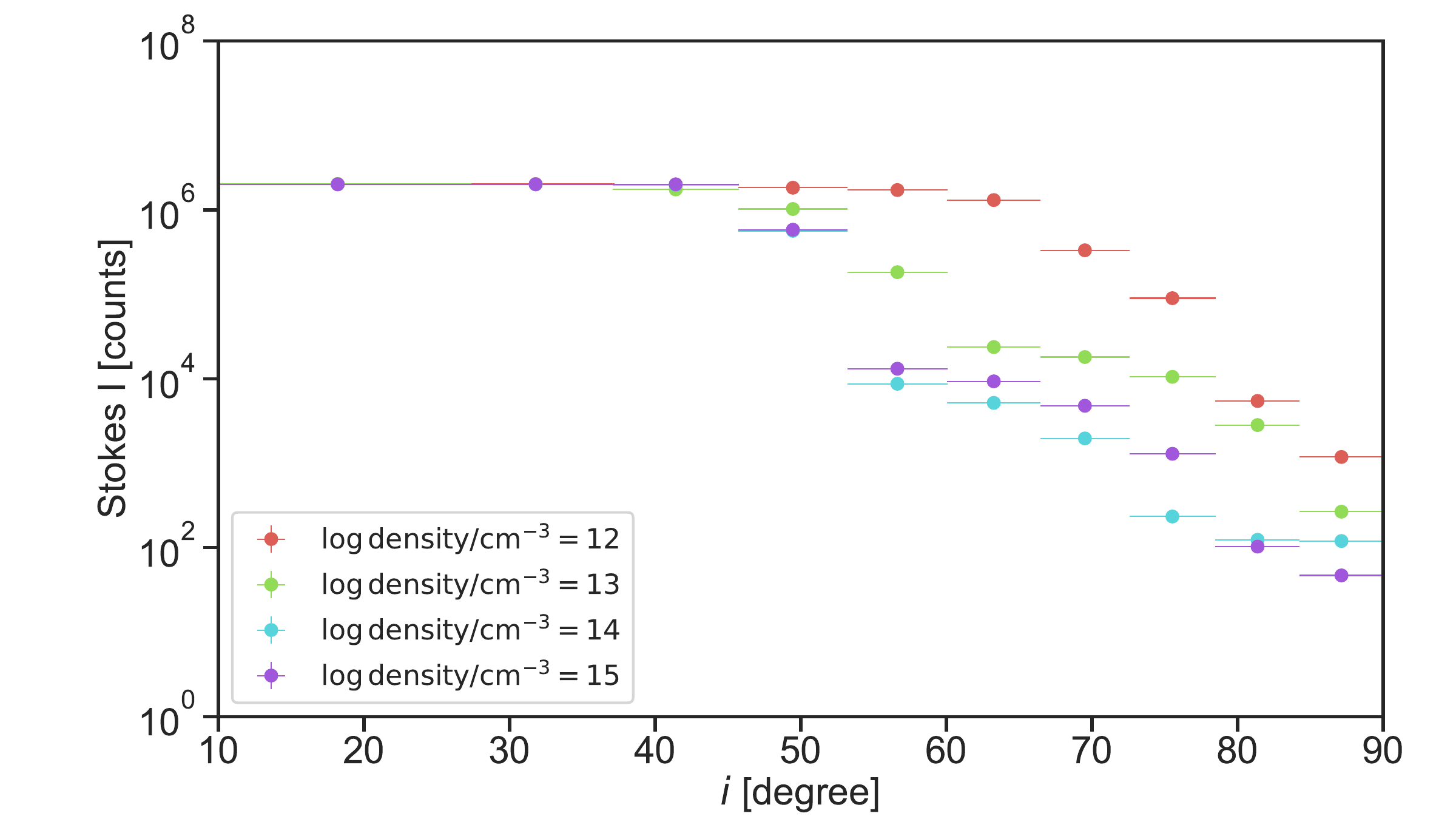}{0.5\textwidth}{(a)}\fig{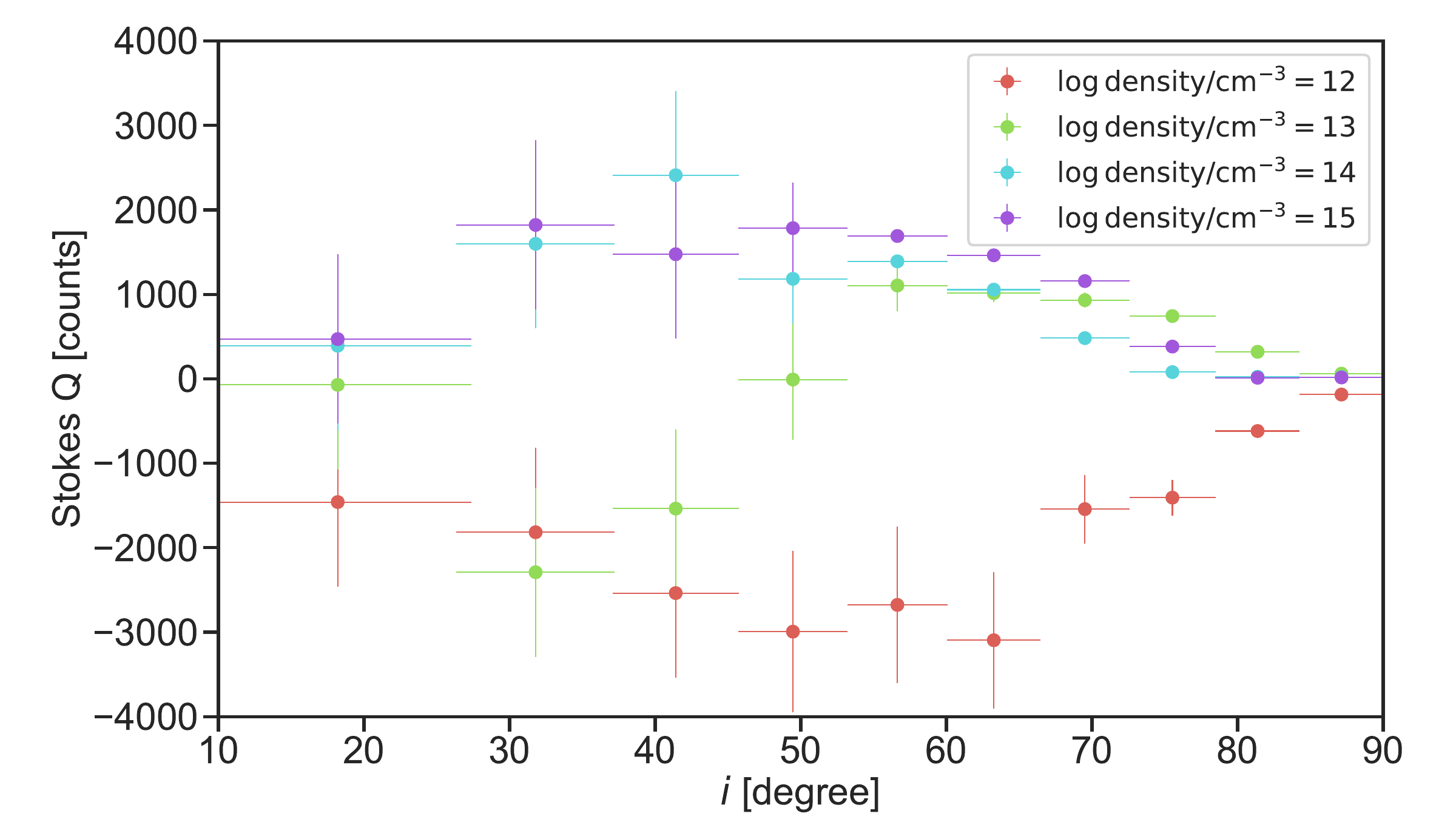}{0.5\textwidth}{(b)}}
\gridline{\fig{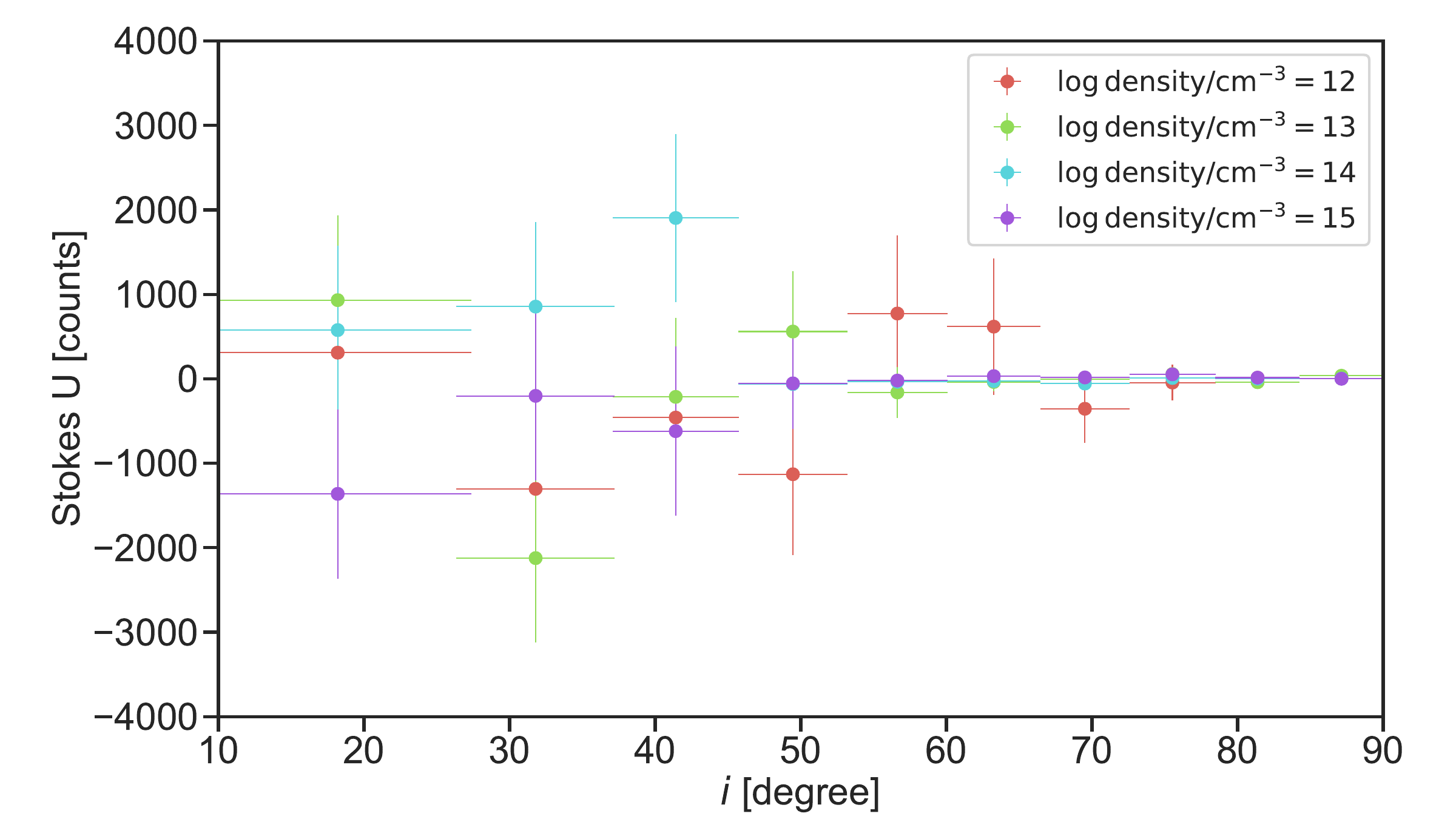}{0.5\textwidth}{(c)}\fig{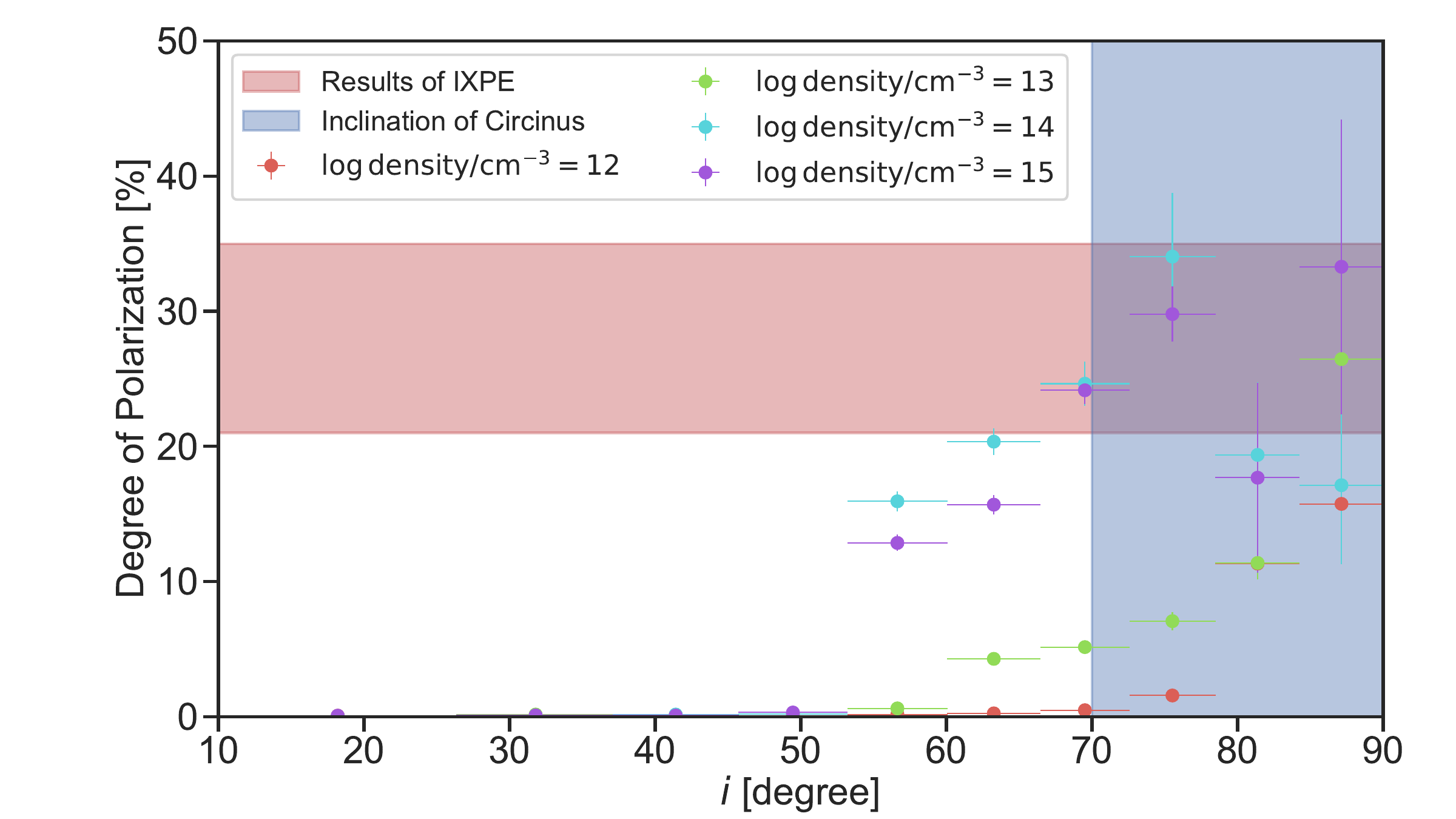}{0.5\textwidth}{(d)}}
\caption{Same as \hyperref[Figure02]{Figure~2}, but for the sub-parsec-scale radiation-driven fountain model.}
\end{figure}

%% file: Section0301.tex
\section{Results}\label{Section0300}
\subsection{Parsec-Scale Radiation-Driven Fountain Model}\label{Section0301}
\hyperref[Figure03]{Figure~3} plots the dependence of (a) the Stokes parameter $I$, (b) the Stokes parameter $Q$, (c) the Stokes parameter $U$, and (d) the degree of polarization on the inclination angle ($i$). Although the IXPE is observable at energies ranging from 2 keV to 8 keV, we integrated these values over the range of 2--6 keV. This is because we avoid the influence of neutral iron X-ray fluorescence at approximately 6.4~keV. In fact, \cite{Ursini2023} removed the polarization contribution from the neutral iron fluorescent line by fixing polconst to zero for the neutral iron fluorescent line. \hyperref[Figure03]{Figure~3(a)} shows that the Stokes parameter $I$ is nearly constant below $i \simeq 50\degr$, whereas it decreases as the inclination angle increase above $i \simeq 50\degr$. This is because the Compton-thin material exists along the line of sight above $i \simeq 50\degr$ (\hyperref[Figure01]{Figure~2(a)}). \hyperref[Figure03]{Figure~3(b)} shows that the Stokes parameter $Q$ is negative. The reason is that, in the parsec-scale radiation-driven fountain model, the material around the nucleus is Compton-thin, and Compton-scattered photons can pass through it. By contrast, if the material around the nucleus is Compton-thick, the Stokes parameter $Q$ will be positive because the photons are Compton-scattered on the surface of the material around the nucleus.

\hyperref[Figure03]{Figure~3(d)} compares the degrees of polarization obtained from our simulation with those of the Circinus galaxy observed with the IXPE \citep{Ursini2023}. \cite{Ursini2023} showed that the degree of polarization for the neutral reflector is $28 \pm 7\%$. It is also known that the inclination angle of the Circinus galaxy is almost edge-on ($i \geq 70\degr$) \citep{Greenhill2003, Tristram2014, Izumi2018, Isbell2022, Ogawa2022}. We found that the degree of polarization obtained from our simulation was smaller than that obtained from observations. This is because the hydrogen column density in this model is small and the polarization effect due to Compton scattering is small. This result is consistent with those of a previous study \citep{Buchner2021}. \cite{Buchner2021} computed the X-ray spectra of neutral materials based on the parsec-scale radiation-driven fountain model, and suggested that a Compton-thick material on the sub-parsec-scale is required.

%% file: Figure05.tex
\begin{figure}\label{Figure05}
\gridline{\fig{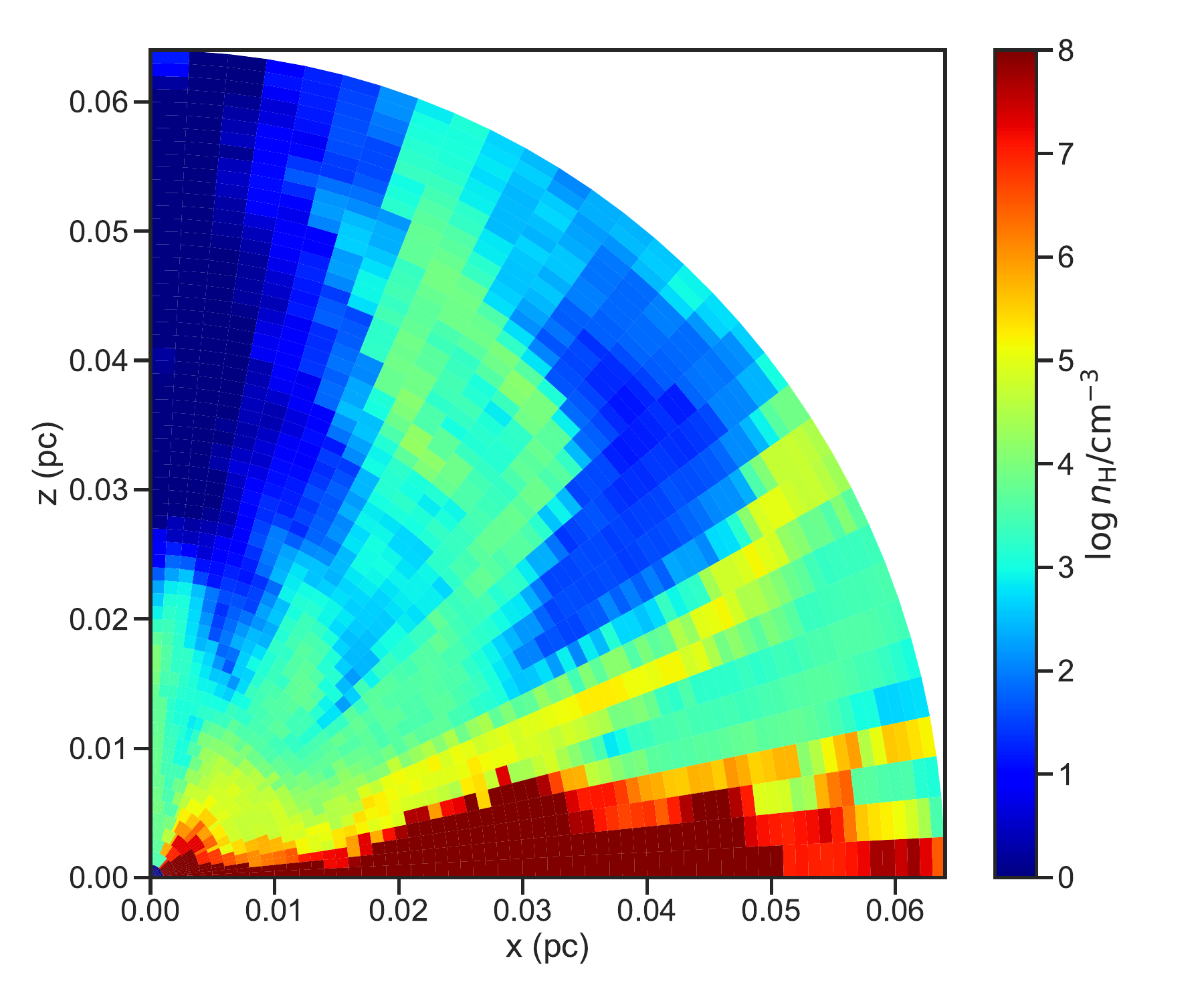}{0.5\textwidth}{(a)}\fig{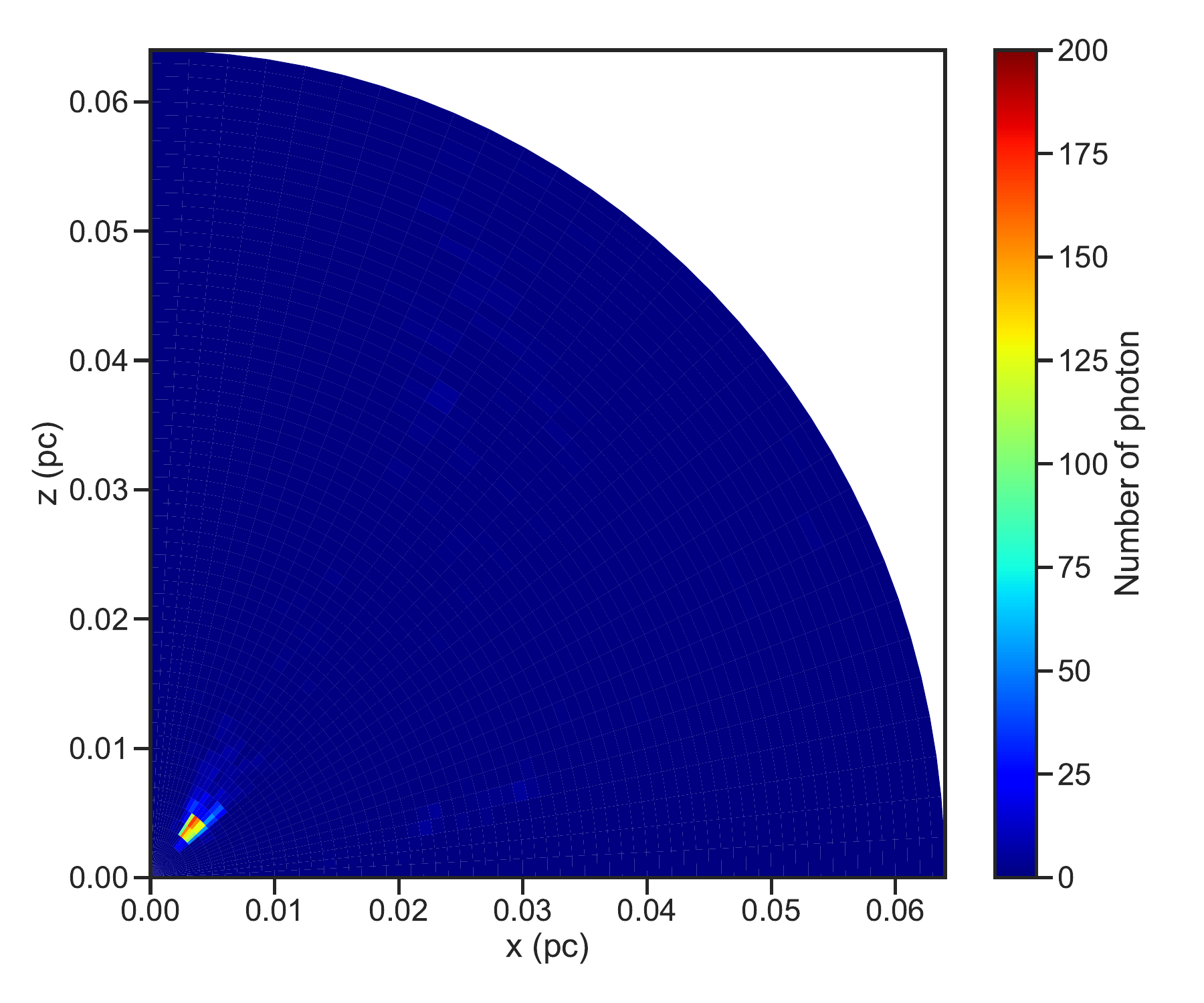}{0.5\textwidth}{(b)}}
\gridline{\fig{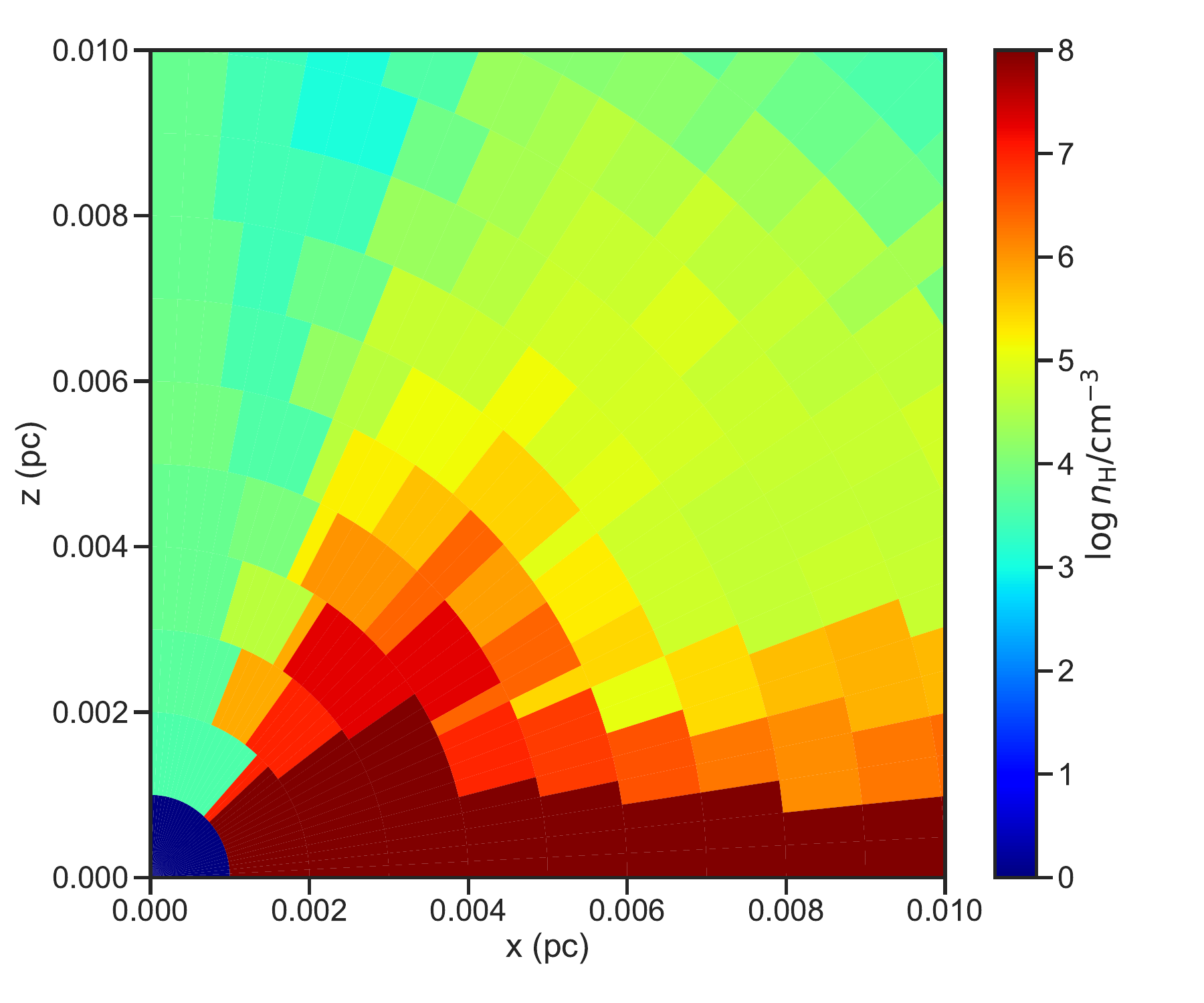}{0.5\textwidth}{(c)}\fig{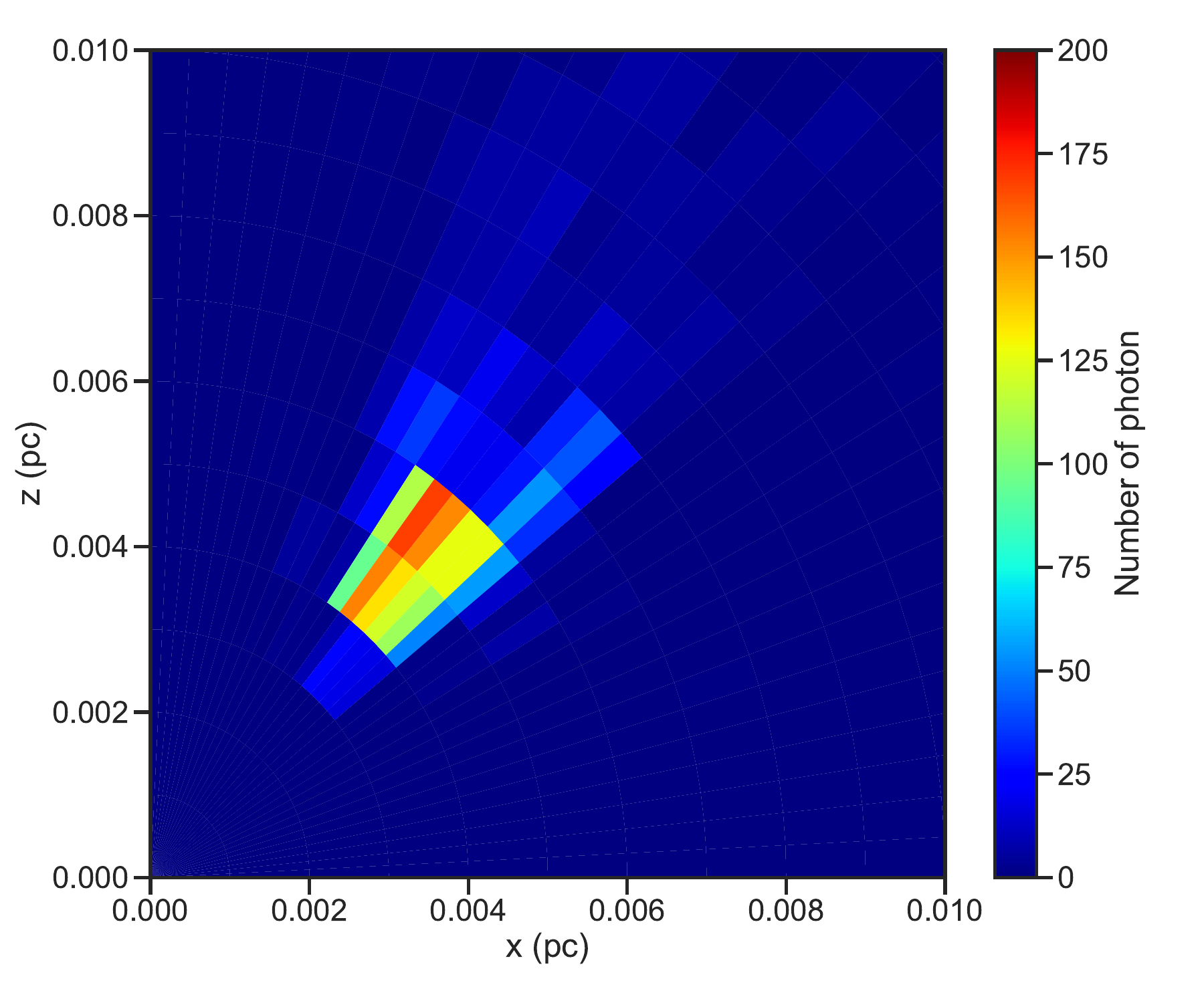}{0.5\textwidth}{(d)}}
\caption{(a) The distribution of the hydrogen number density ($\log n_{\mathrm{H}}/\mathrm{cm}^{-3}$) inside $0.064 \ \mathrm{pc}$. (b) The distribution of the number of Compton-scattered photons in the case of the edge-on view inside $0.064 \ \mathrm{pc}$. (c) The distribution of $\log n_{\mathrm{H}}/\mathrm{cm}^{-3}$ inside $0.01 \ \mathrm{pc}$. (d) The distribution of the number of Compton-scattered photons inside $0.01 \ \mathrm{pc}$. Here we assume the hydrogen number density of the disk of $\log d/\mathrm{cm}^{-3} = 15$.}
\end{figure}

%% file: Figure06.tex
\begin{figure}\label{Figure06}
\gridline{\fig{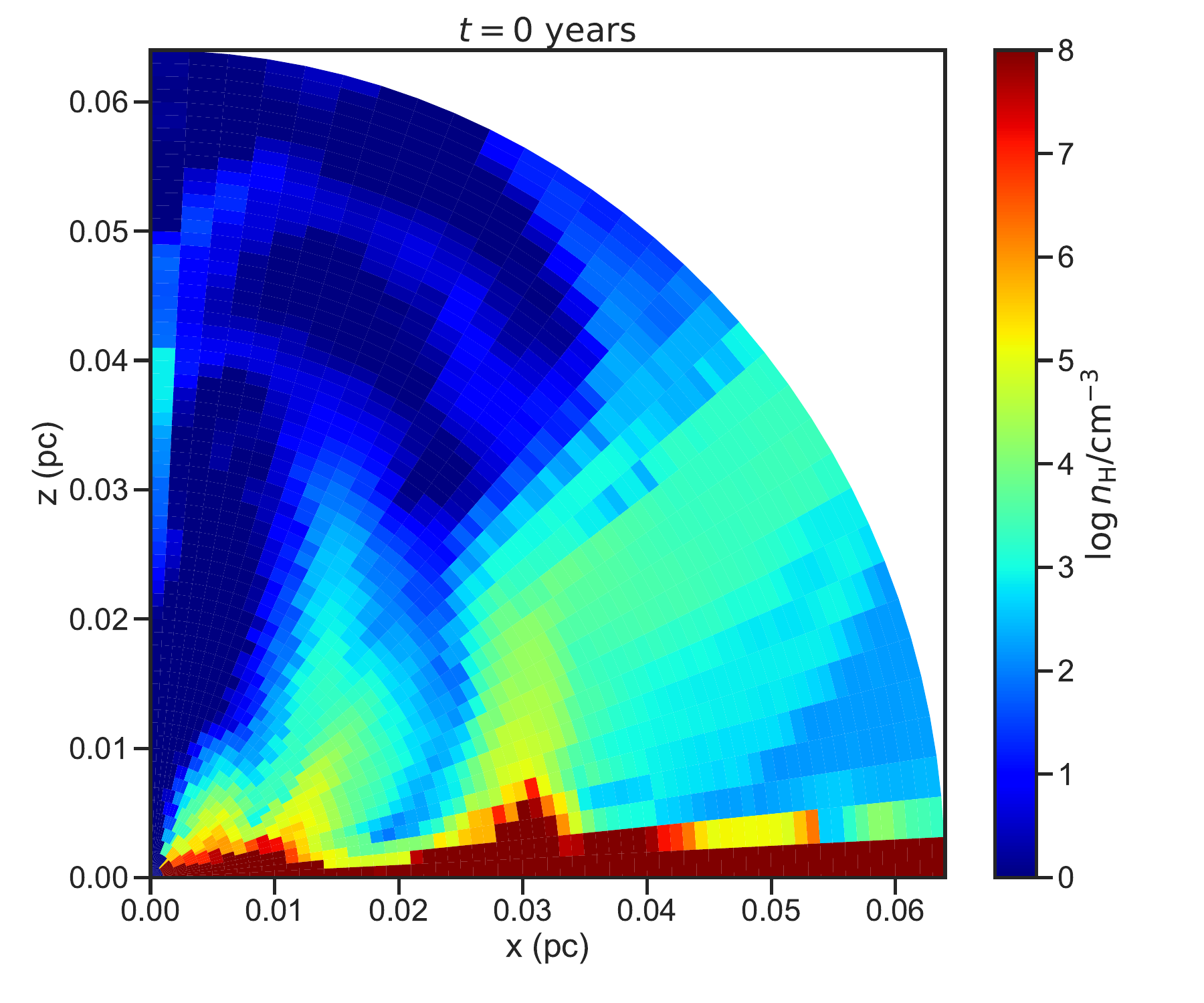}{0.5\textwidth}{(a)}\fig{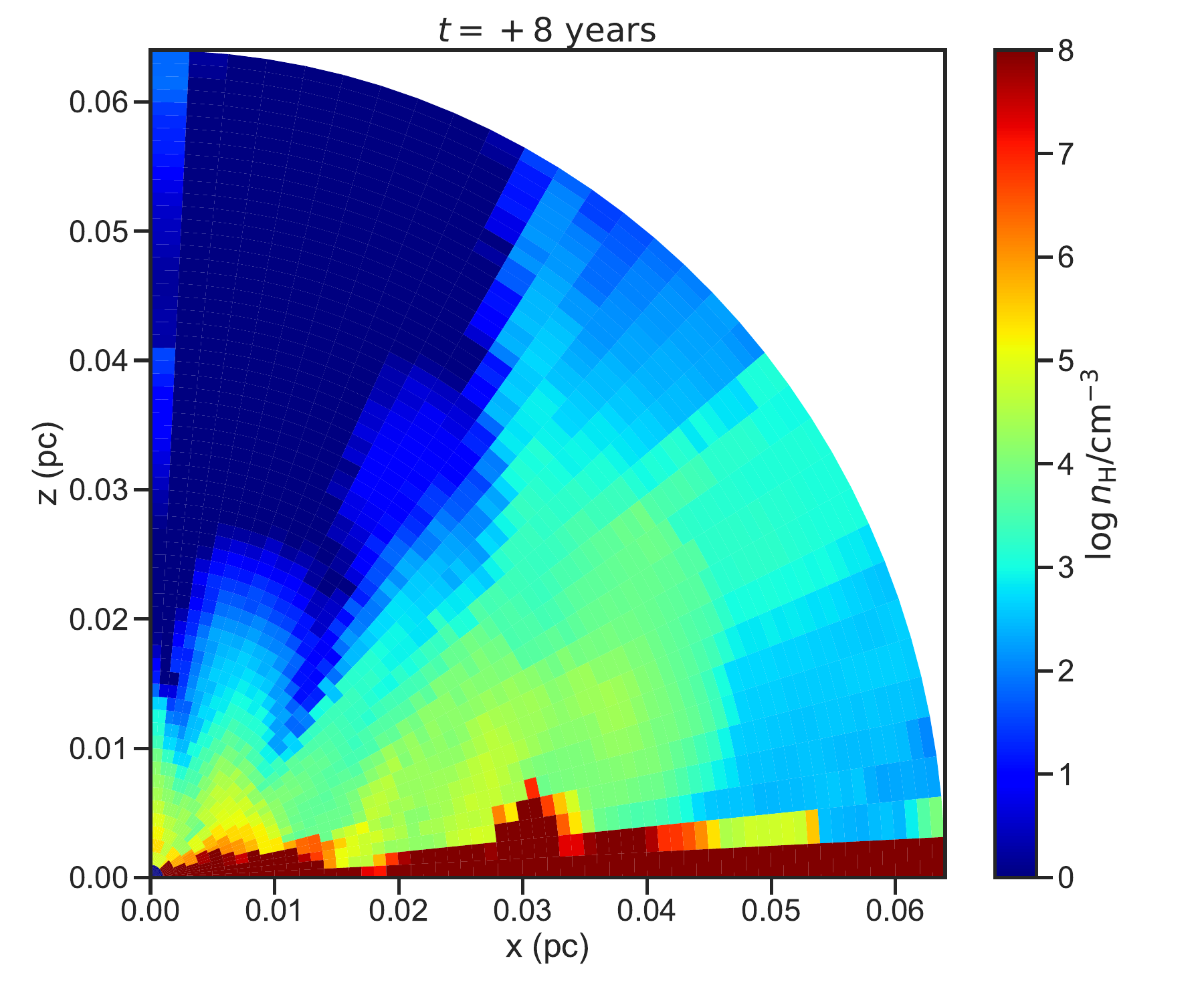}{0.5\textwidth}{(b)}}
\gridline{\fig{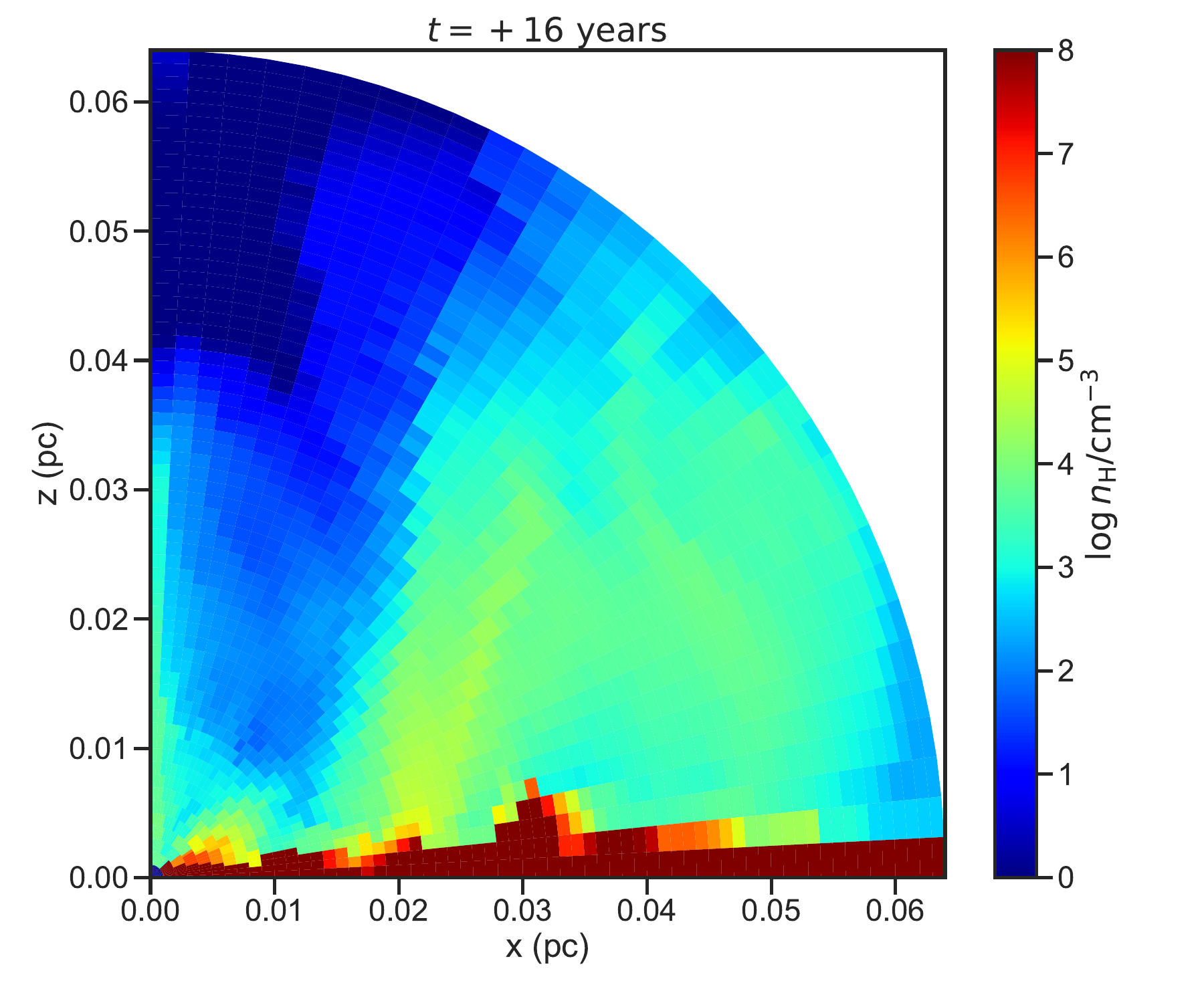}{0.5\textwidth}{(c)}\fig{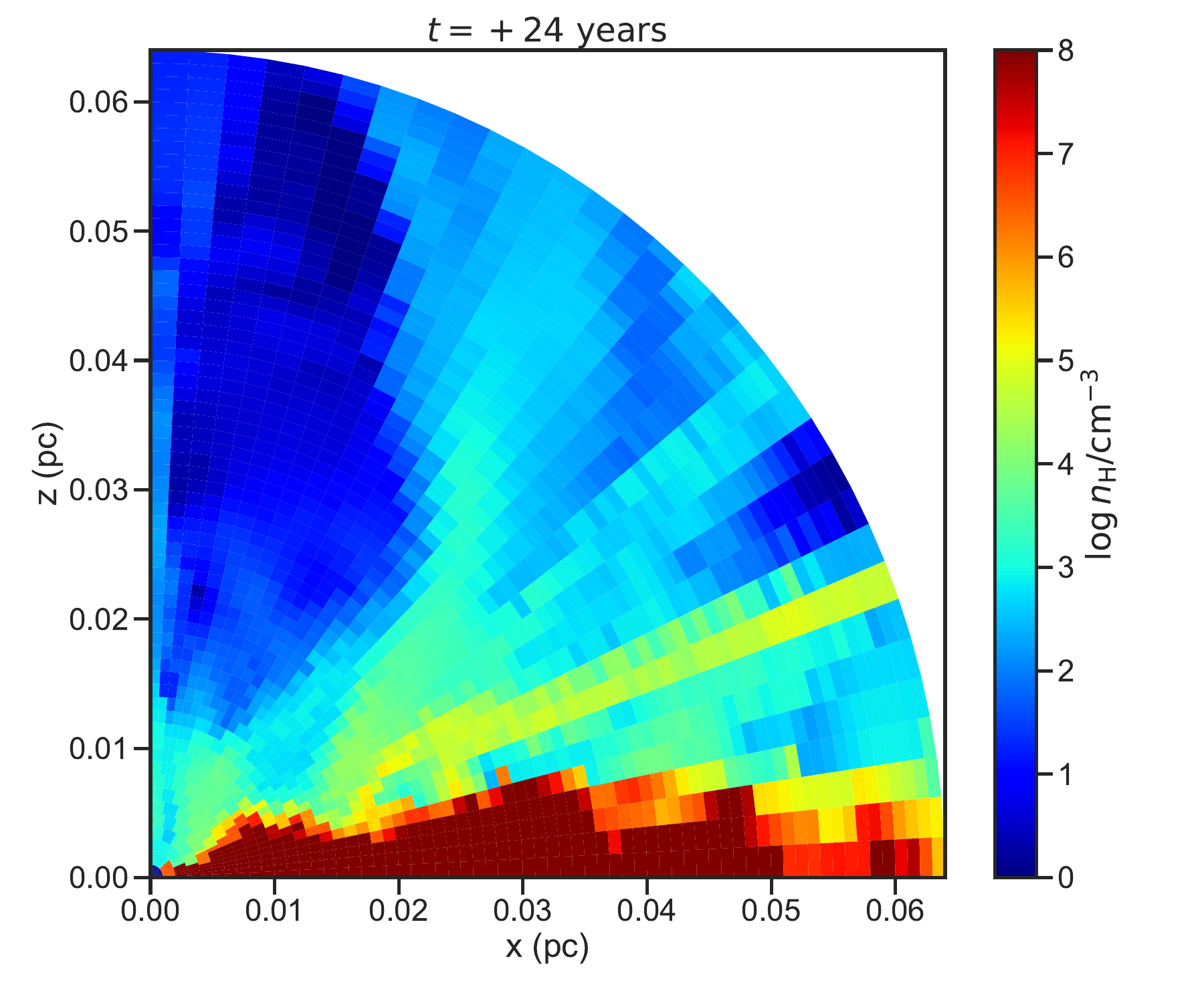}{0.5\textwidth}{(d)}}
\caption{(a) The distribution of the hydrogen number density ($\log n_{\mathrm{H}}/\mathrm{cm}^{-3}$) based on the sub-parsec-scale radiation-driven fountain model inside $0.064 \ \mathrm{pc}$. (b) The distribution of $\log n_{\mathrm{H}}/\mathrm{cm}^{-3}$ after eight years. (c) The distribution of $\log n_{\mathrm{H}}/\mathrm{cm}^{-3}$ after 16 years. (d) The distribution of $\log n_{\mathrm{H}}/\mathrm{cm}^{-3}$ after 24 years. Here we assume the hydrogen number density of the disk of $\log d/\mathrm{cm}^{-3} = 14$.}
\end{figure}

%% file: Section0302.tex
\subsection{Sub-Parsec-Scale Radiation-Driven Fountain Model}\label{Section0302}
We then computed the Stokes parameters based on the sub-parsec-scale radiation-driven fountain model \citep{Kudoh2023}. \hyperref[Figure04]{Figure~4} plots the dependence of (a) the Stokes parameter $I$, (b) the Stokes parameter $Q$, (c) the Stokes parameter $U$, and (d) the degrees of polarization on the hydrogen number density of the disk ($d$) and the inclination angle ($i$). Here we integrated the values from 2~keV to 6~keV. \hyperref[Figure04]{Figure~4(a)} shows that the Stokes parameter $I$ decrease with $d$ and $i$. \hyperref[Figure04]{Figure~4(b)} shows that the Stokes parameter $Q$ is negative in the case of $i \geq 70\degr$ and $\log d/\mathrm{cm}^{-3} = 12$, whereas the Stokes parameter $Q$ is positive in the case of $i \geq 70\degr$ and $\log d/\mathrm{cm}^{-3} \geq 13$. As explained in \hyperref[Section0301]{Section~3.1}, this is because the sub-parsec-scale radiation-driven fountain model has a Compton-thick material in the case of $i \geq 70\degr$ (\hyperref[Figure02]{Figure~2(b)}). \hyperref[Figure04]{Figure~4(c)} shows that the Stokes parameter $U$ is almost constant in the case of $i \geq 70\degr$ and $\log d/\mathrm{cm}^{-3} \geq 13$.

\hyperref[Figure04]{Figure~4(d)} compares the degrees of polarization obtained from our simulation with those of the Circinus galaxy. In the case of $i \geq 70\degr$ and $\log d/\mathrm{cm}^{-3} \geq 13$, we found that the degree of polarization obtained from our simulation was consistent with that inferred from the IXPE observation. We also compare the polarization angle obtained from our simulation with that observed with the IXPE. The polarization angle obtained from the IXPE observation is $\psi = 18\pm5\degr$ \citep{Ursini2023}. This value is close to the direction of the inner $\mathrm{H_{2}O}$ maser disk ($29\degr$; \citealt{Greenhill2003}) and is consistent with being perpendicular to the radio jet ($295\degr$; \citealt{Elmouttie1998a}) and to the axis of the $\mathrm{H}\alpha$ ionization cone \citep{Elmouttie1998b}. As explained in \hyperref[Section0202]{Section 2.2}, we cannot compare the absolute values of the polarization angles from our calculations with those from the IXPE observations, but the value of the polarization angle from our calculations is almost $0\degr$, which is consistent with the disk direction. 

%% file: Figure07.tex
\begin{figure}\label{Figure07}
\gridline{\fig{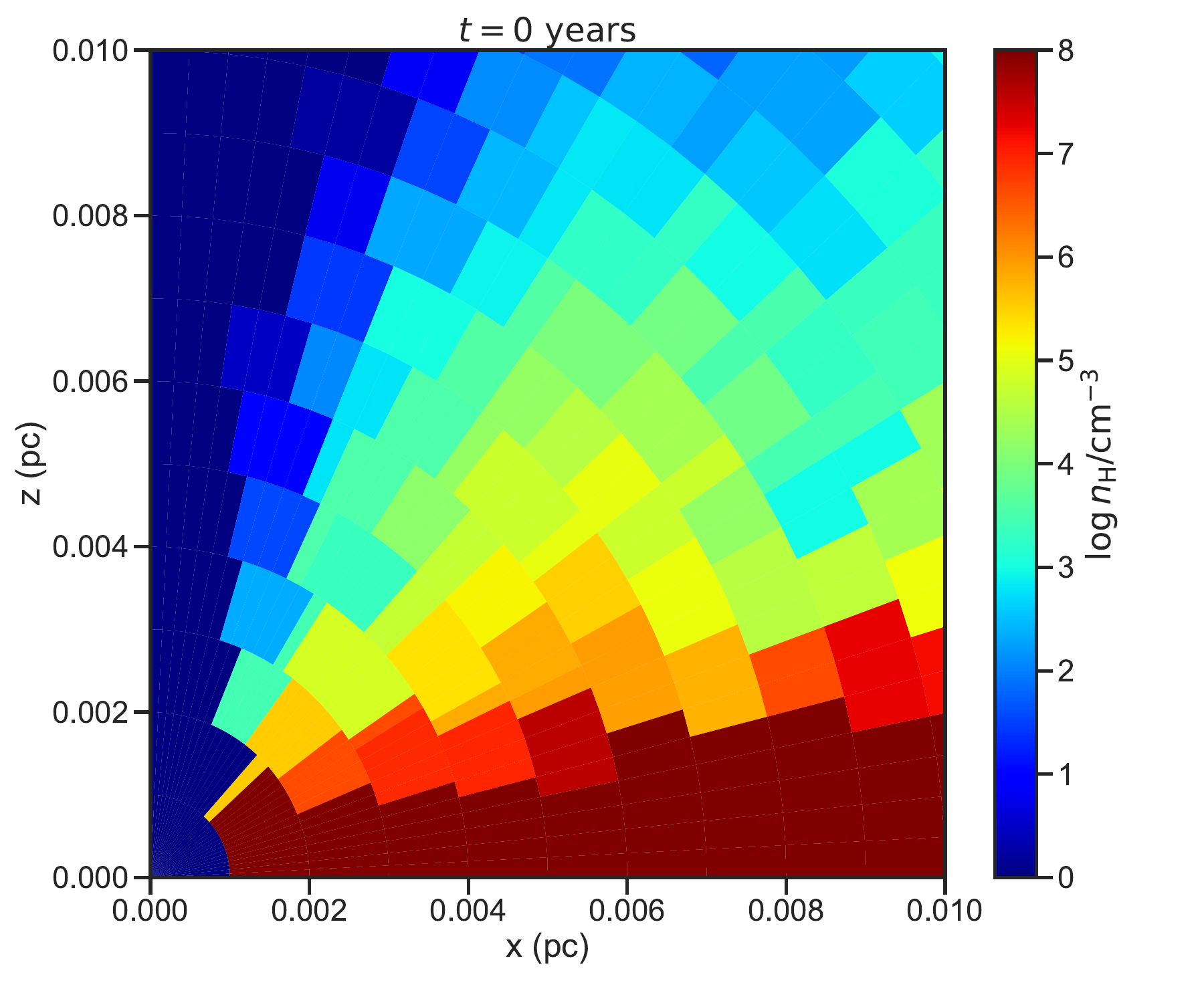}{0.5\textwidth}{(a)}\fig{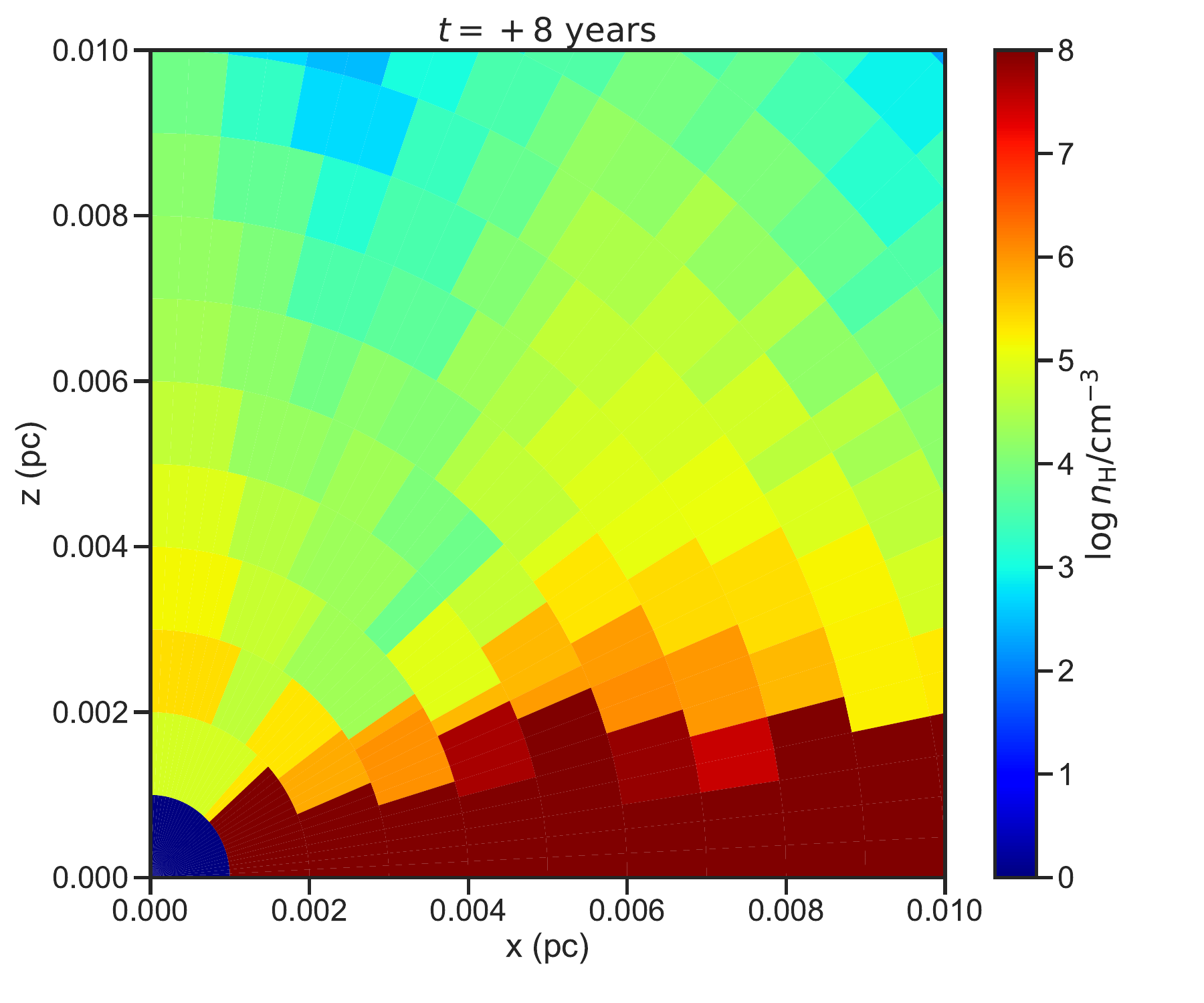}{0.5\textwidth}{(b)}}
\gridline{\fig{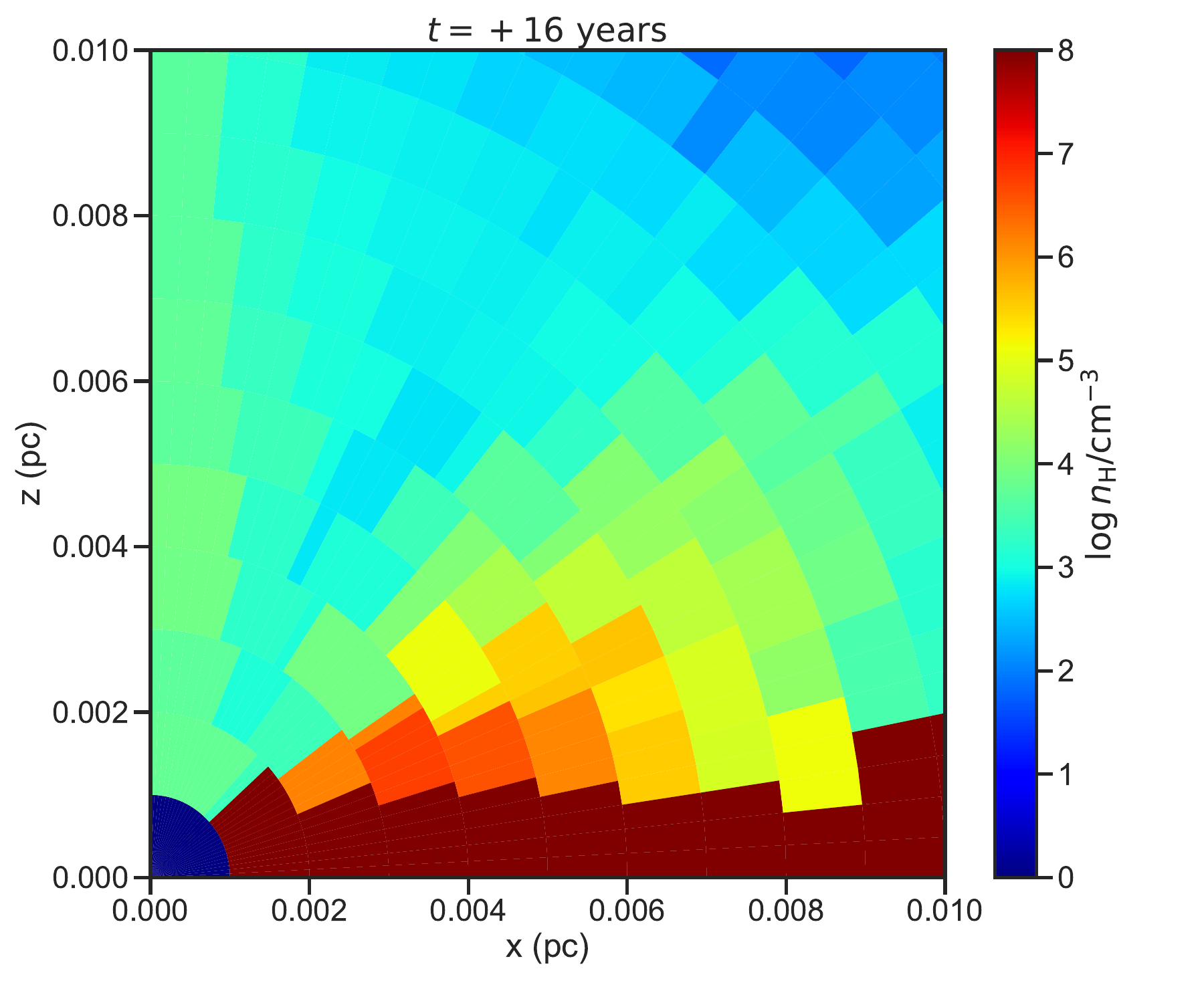}{0.5\textwidth}{(c)}\fig{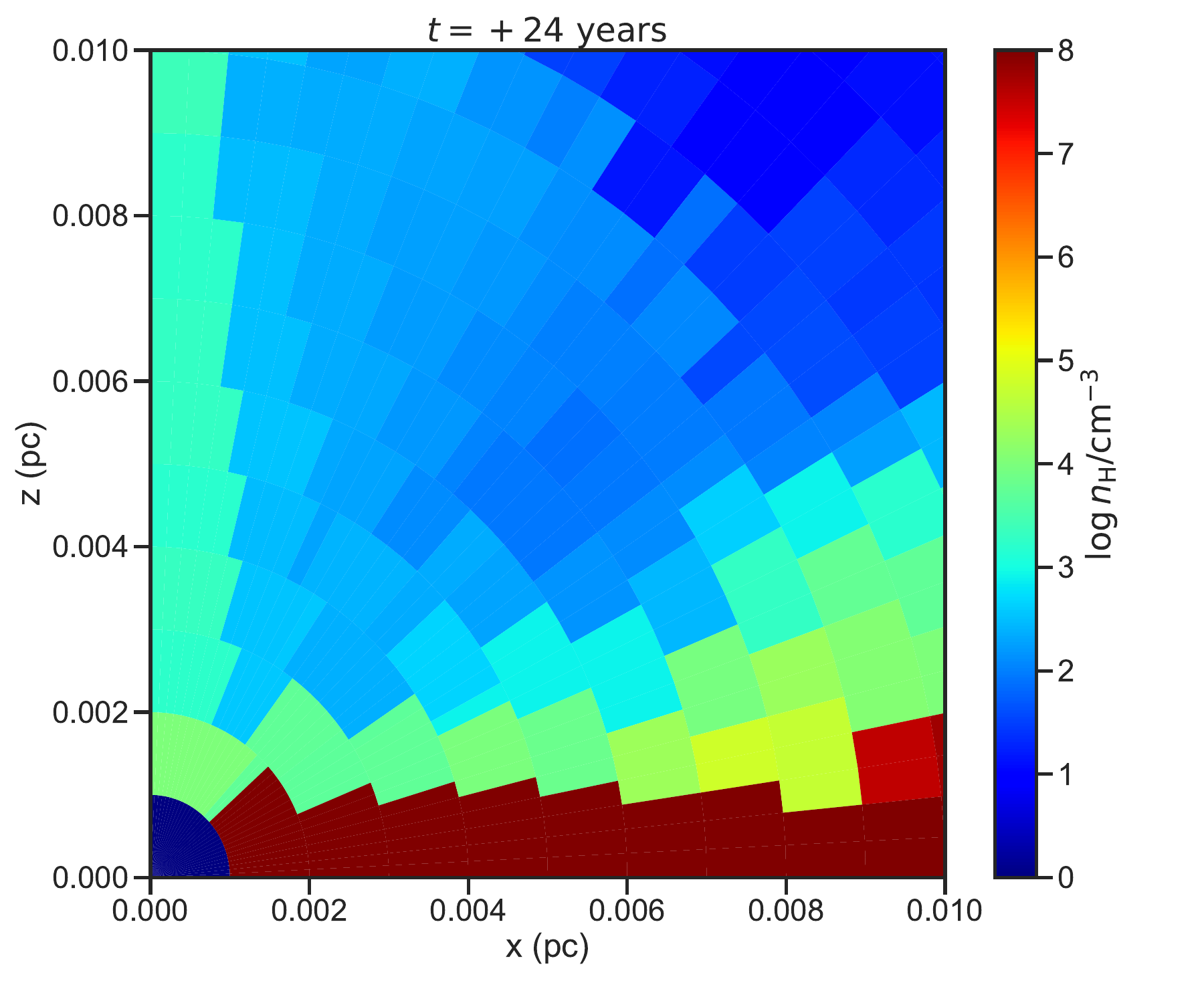}{0.5\textwidth}{(d)}}
\caption{(a) The distribution of the hydrogen number density ($\log n_{\mathrm{H}}/\mathrm{cm}^{-3}$) based on the sub-parsec-scale radiation-driven fountain model inside $0.01 \ \mathrm{pc}$. (b) The distribution of $\log n_{\mathrm{H}}/\mathrm{cm}^{-3}$ after eight years. (c) The distribution of $\log n_{\mathrm{H}}/\mathrm{cm}^{-3}$ after 16 years. (d) The distribution of $\log n_{\mathrm{H}}/\mathrm{cm}^{-3}$ after 24 years. Here we assume the hydrogen number density of the disk of $\log d/\mathrm{cm}^{-3} = 14$.}
\end{figure}

%% file: Figure08.tex
\begin{figure}\label{Figure08}
\gridline{\fig{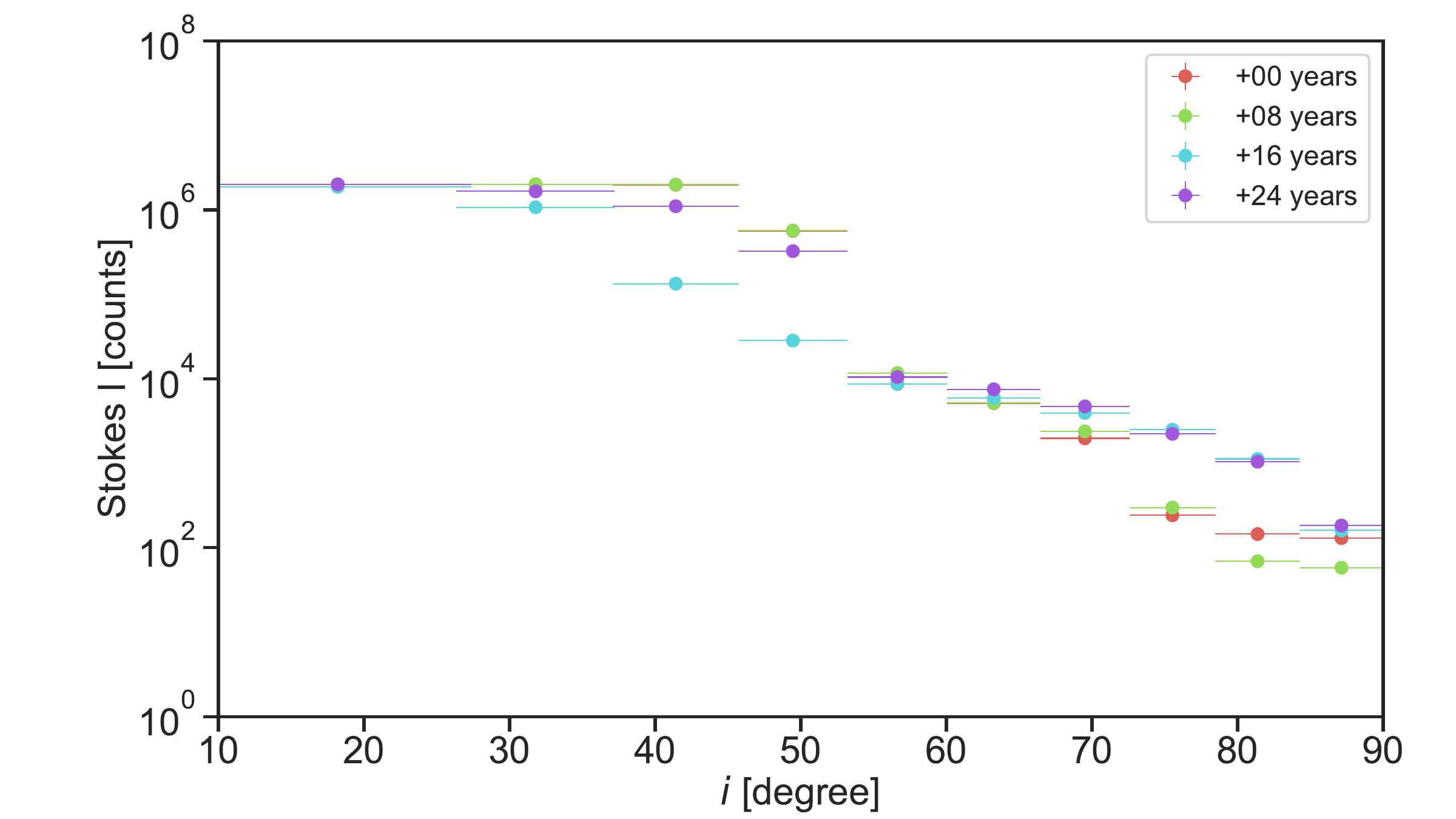}{0.5\textwidth}{(a)}\fig{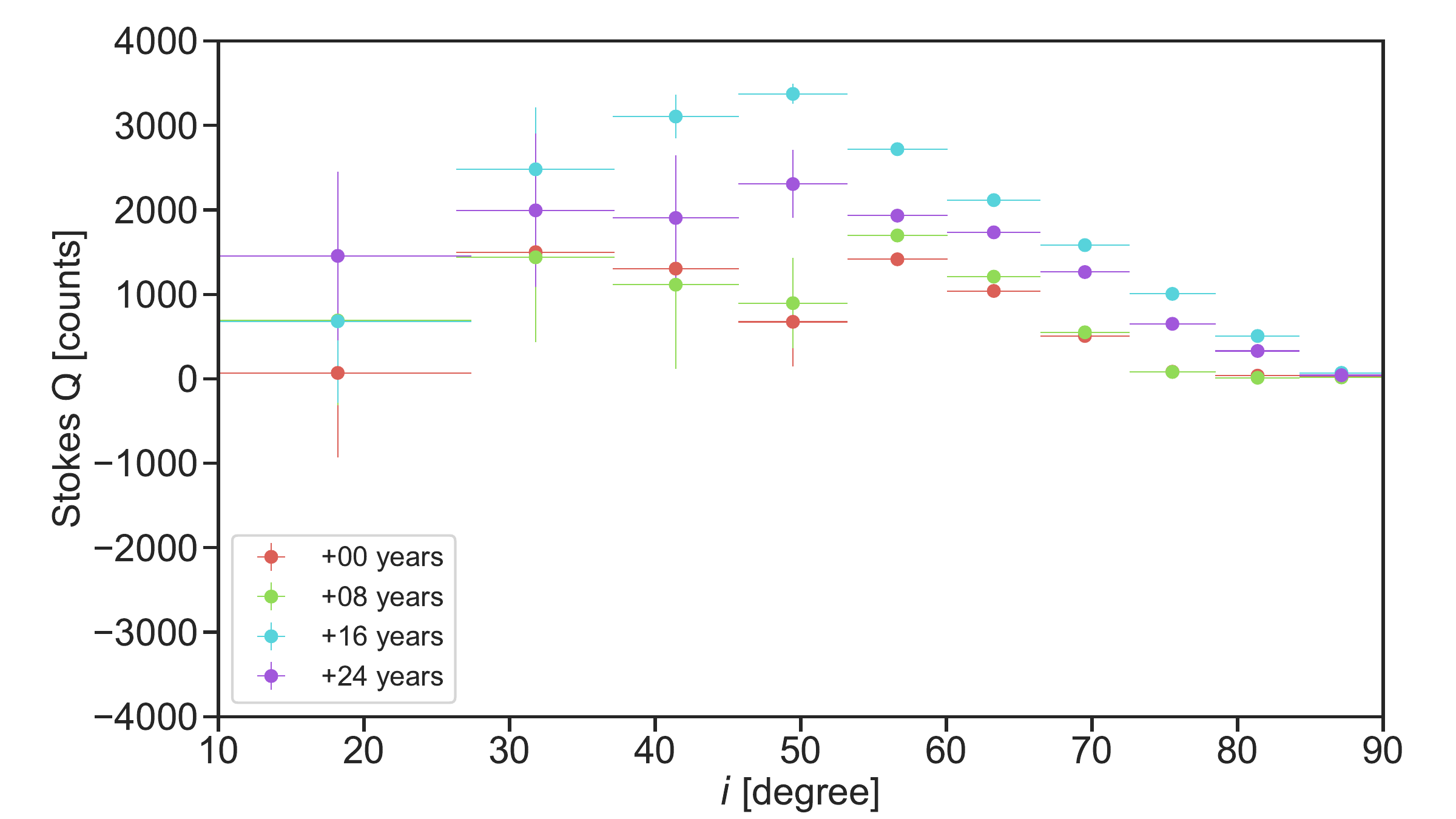}{0.5\textwidth}{(b)}}
\gridline{\fig{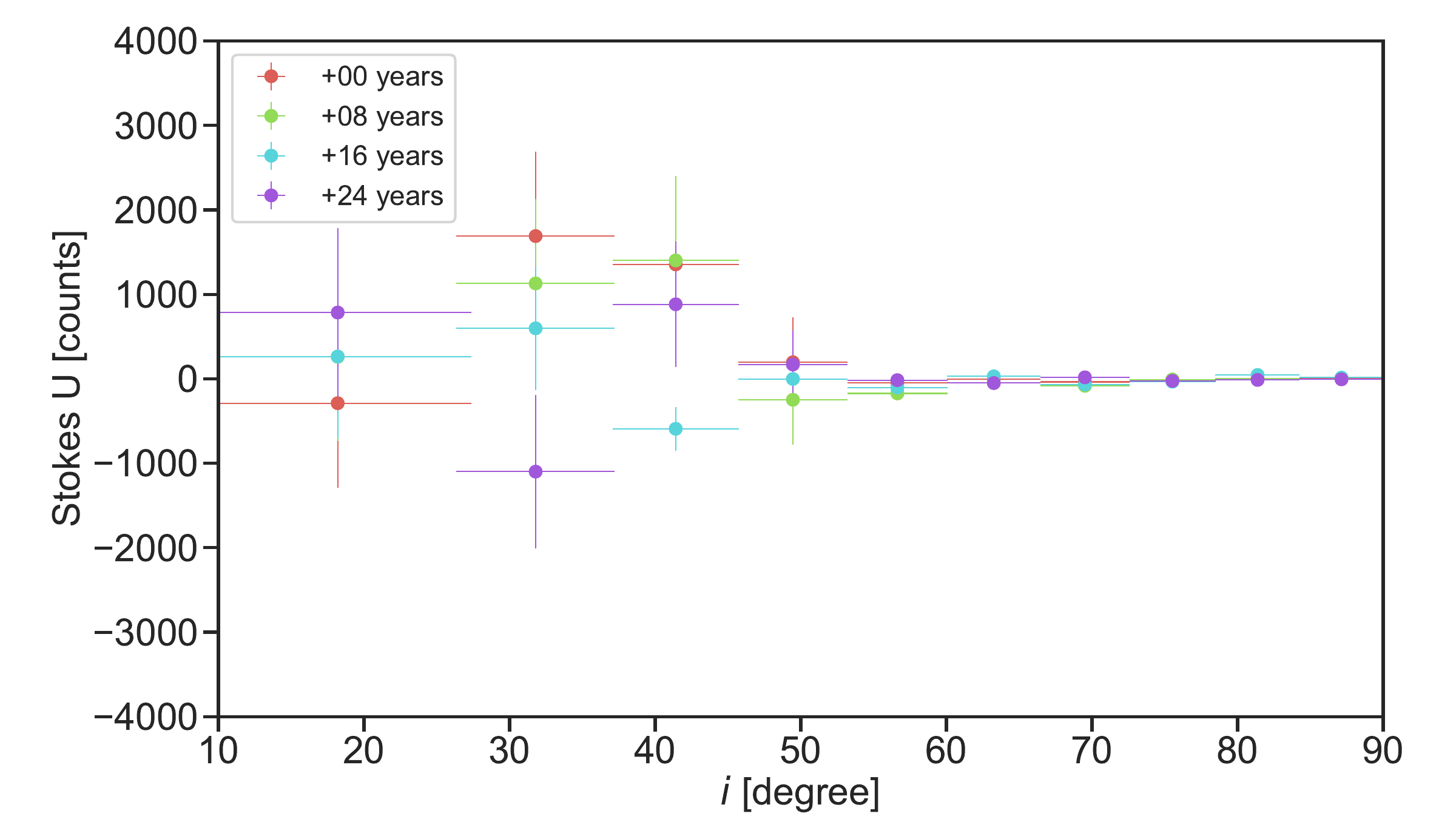}{0.5\textwidth}{(c)}\fig{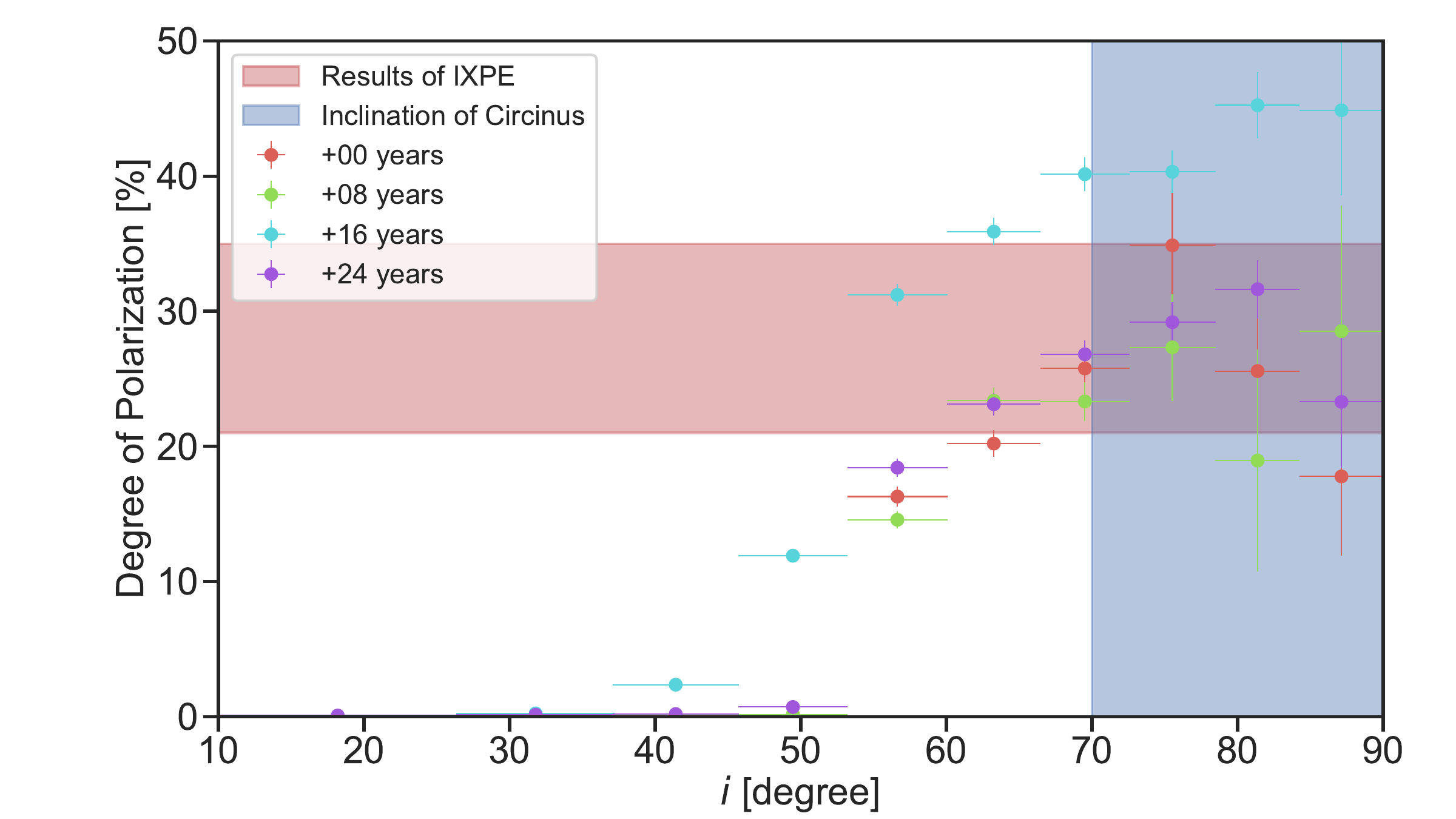}{0.5\textwidth}{(d)}}
\caption{(a) The dependence of the Stokes parameter $I$ on the time. (b) The dependence of the Stokes parameter $Q$ on the time. (c) The dependence of the Stokes parameter $U$ on the time. (d) Comparison of the degree of polarization obtained from our simulation with that of the Circinus galaxy observed with the IXPE (red region). The blue region represents a range of observational suggested inclination angles of the Circinus galaxy (see \hyperref[Section0301]{Section~3.1}). Here errors correspond to a 1$\sigma$ credible interval.}
\end{figure}

%% file: Section0401.tex
\section{Discussion}\label{Section0400}
\subsection{Origin of X-Ray Polarization}\label{Section0401}
To determine the origin of the X-ray polarization in the Circinus galaxy, we investigated where the photons were Compton-scattered. \hyperref[Figure05]{Figure~5} plots (a) the distribution of the hydrogen number density ($\log n_{\mathrm{H}}/\mathrm{cm}^{-3}$) inside $0.064 \ \mathrm{pc}$, (b) the distribution of the number of Compton-scattered photons in the case of the edge-on view inside $0.064 \ \mathrm{pc}$, (c) the distribution of $\log n_{\mathrm{H}}/\mathrm{cm}^{-3}$ inside $0.01 \ \mathrm{pc}$, and (d) the distribution of the number of Compton-scattered photons inside $0.01 \ \mathrm{pc}$. Although we assumed the hydrogen number density of the disk of $\log d/\mathrm{cm}^{-3} = 15$, we confirmed that the trend was the same for different values. \hyperref[Figure05]{Figure~5(b)} suggests that almost all photons are Compton scattered inside $0.01 \ \mathrm{pc}$. In particular, the photons are mainly Compton scattered approximately $r \simeq 0.005 \ \mathrm{pc}$ whose hydrogen number density is approximately $10^{7} \ \mathrm{cm}^{-3}$ (\hyperref[Figure05]{Figure~5(c)-(d)}). This implies that the origin of X-ray polarization in the Circinus galaxy is a sub-parsec-scale radiation-driven outflow.

%% file: Section0402.tex
\subsection{Time Variability of Degree of Polarization}\label{Section0402}
We examined the time variability in the degree of polarization. This is because the structures of the dense gas in the central region that contribute to the polarization depend on time. \hyperref[Figure06]{Figure~6} plots (a) the distribution of the hydrogen number density ($\log n_{\mathrm{H}}/\mathrm{cm}^{-3}$) based on the sub-parsec-scale radiation-driven fountain model inside $0.064 \ \mathrm{pc}$, (b) the distribution of $\log n_{\mathrm{H}}/\mathrm{cm}^{-3}$ after eight years, (c) the distribution of $\log n_{\mathrm{H}}/\mathrm{cm}^{-3}$ after 16 years, and (d) the distribution of $\log n_{\mathrm{H}}/\mathrm{cm}^{-3}$ after 24 years. Here we assume the hydrogen number density of the disk to be $\log d/\mathrm{cm}^{-3} = 14$. \hyperref[Figure07]{Figure~7} also plots the distribution of $\log n_{\mathrm{H}}/\mathrm{cm}^{-3}$ inside $0.01 \ \mathrm{pc}$. \hyperref[Figure07]{Figure~7} suggests that the distribution of $\log n_{\mathrm{H}}/\mathrm{cm}^{-3}$ inside $0.01 \ \mathrm{pc}$, which is thought to be the origin of X-ray polarization, changes on a timescale of approximately $10$ years. This can affect the degree of polarization.

\hyperref[Figure08]{Figure~8} plots the dependence of (a) the Stokes parameter $I$, (b) the Stokes parameter $Q$, (c) the Stokes parameter $U$, and (d) the degrees of polarization on time. \hyperref[Figure08]{Figures~8(a)--(c)} show that the Stokes parameters $I$, $Q$, and $U$ depend on time. In particular, the Stokes parameters $I$ and $Q$ strongly depend on time. \hyperref[Figure08]{Figure~8(d)} shows that the degree of polarization varies in the range of $20\%$--$45\%$. This implies that monitoring observations of X-ray polarization over timescales of a few years may be important to understand the origin of X-ray polarization. We also studied the time variability in the polarization angle and found that the polarization angles agreed within an error. Our simulations are based on two-dimensional axisymmetric radiative hydrodynamic simulation results, which may overestimate the time variability in the degree of polarization.

%% file: Section0501.tex
\section{Conclusion}\label{Section0500}
We computed the degree of polarization based on two types of radiative hydrodynamic simulations: the parsec-scale radiation-driven fountain model \citep{Wada2016}, and the sub-parsec-scale radiation-driven fountain model \citep{Kudoh2023} using MONACO code \citep{Odaka2011, Odaka2016} version 1.7.s1. The following conclusions were drawn.

\begin{enumerate}
\item We compared the degree of polarization based on the parsec-scale radiation-driven fountain model with that of the Circinus galaxy observed with the IXPE. The degree of polarization obtained from our simulation was smaller than that inferred from the IXPE observations.
\item We studied the dependence of the degree of polarization on the hydrogen number density of the disk ($\log d/\mathrm{cm}^{-3}$) based on the sub-parsec-scale radiation-driven fountain model. We found that the degree of polarization depends on $\log d/\mathrm{cm}^{-3}$. In the cases of $\log d/\mathrm{cm}^{-3} \geq 13$ and $i \geq 70\degr$, the degree of polarization obtained from our simulation was consistent with that inferred from the IXPE observations.
\item We investigated where the photons are Compton scattered and implied that the origin of the X-ray polarization in the Circinus galaxy is the outflow inside $0.01 \ \mathrm{pc}$.
\item We evaluated the time variability in the degree of polarization. We found that the degree of polarization changed over approximately $10$ years. If the time variability of the degree of polarization is detected in the Circinus galaxy in the future, it reinforces that the origin of the X-ray polarization is the gas inside $0.01 \ \mathrm{pc}$.
\end{enumerate}

%% file: Acknowledgment.tex
\begin{acknowledgements}
We thank the anonymous referee for the helpful comments. Atsushi Tanimoto and the present research are partly supported by the Kagoshima University postdoctoral research program (KU-DREAM). This work is also supported by the Grant-in-Aid for JSPS Fellows (R.U.) and the Grants-in-Aid for Scientific Research 22H00128 (H.O.). Numerical computations were performed on Cray XC50 at the Center for Computational Astrophysics, National Astronomical Observatory of Japan.
\end{acknowledgements}
\facilities{IXPE \citep{Weisskopf2022}.}
\software{CANS+ \citep{Matsumoto2019}, MONACO \citep{Odaka2011, Odaka2016}.}